\documentclass[journal,10pt]{IEEEtran}
\usepackage{float}
\usepackage{comment}
\usepackage[table]{xcolor}
\usepackage{threeparttable,booktabs}
\usepackage{cite}      
\usepackage{graphicx}  
\usepackage{psfrag}    
\usepackage{placeins}
\usepackage{lineno}
\usepackage{url}
\usepackage{multirow}
\usepackage{subcaption}
\usepackage{caption}
\usepackage{times}
\usepackage{textcomp}
\usepackage{amsmath} 
\usepackage{amsfonts,amsthm,amssymb,mathtools}
\usepackage{balance}
\usepackage[normalem]{ulem}
\usepackage{xcolor}

\usepackage{cuted}
\usepackage{amsmath,bm}
\usepackage{hyperref}
\usepackage{cleveref}
\usepackage{nccmath}
\usepackage{tabularx} 

\usepackage{array} 
\newcolumntype{P}[1]{>{\centering\arraybackslash}p{#1}}

\begin{document}
\title{Radar Sensing Based on 1-Bit Quantized Reconfigurable Intelligent Surfaces}
\author{\IEEEauthorblockN{Kainat~Yasmeen,  Shobha~Sundar~Ram, Debidas Kundu}

\thanks{K. Yasmeen is with Indraprastha Institute of Information Technology Delhi (email: kainaty@iiitd.ac.in); S. S. Ram is currently affiliated with University College London, (s.ram@ucl.ac.uk). D. Kundu was with Indian Institute of Technology Delhi. He is currently with Indian Institute of Technology Bombay (email: debidask@ee.iitb.ac.in).}
\thanks{This manuscript has been accepted for publication in
IEEE Transactions on Radar Systems.
The final version of record is forthcoming.}}
\maketitle

\begin{abstract}
We present a radar sensing framework based on a low-complexity, quantized reconfigurable intelligent surface (RIS) that enables programmable manipulation of electromagnetic wavefronts for enhanced detection in non-specular and shadowed regions. We develop closed-form expressions for the scattered field and radar cross section (RCS) of phase-quantized RIS apertures based on aperture field theory hybridized with full-wave simulations, accurately capturing the effects of quantized phase, periodicity, and grating lobes on radar detection performance. The theory enables us to analyze the RIS's RCS along both the forward and backward paths from the radar to the target. The theory is benchmarked against full-wave electromagnetic simulations incorporating realistic unit-cell amplitude and phase responses. To validate practical feasibility, a $[16\times10]$ 1-bit RIS operating at 5.5 GHz is fabricated and experimentally characterized inside an anechoic chamber and in real radio channels. Measurements of steering angles, beam-squint errors, and peak-to-specular ratios of the RCS patterns exhibit strong agreement with analytical and simulated results. Further experiments demonstrate that the RIS can redirect the beam in a non-specular direction and recover micro-Doppler signatures that remain undetectable with a conventional radar deployment. 
\end{abstract}

\providecommand{\keywords}[1]{\textbf{\textit{Keywords--}}#1}

\section{Introduction}
\label{sec:Introduction}
Reconfigurable intelligent surfaces (RIS) have emerged as a powerful approach for programmable control of electromagnetic wave propagation. By using large arrays of tunable subwavelength elements, RIS can control the phase of incoming waves and steer the reflected field in desired directions \cite{ozdogan2019intelligent,liu2024efficient,yu2022reconfigurable}. While RISs have been widely studied for enhancing wireless communication links \cite{huang2019reconfigurable,10715713,wu2021intelligent,elmossallamy2020reconfigurable,liu2021reconfigurable}, their use in radar and electromagnetic scattering applications has received comparatively limited attention. Radar sensing in cluttered environments with obstacles is often challenged by the lack of reliable line-of-sight (LOS) paths \cite{ram2008through,ram2010simulation}. These challenges become more severe at higher operating frequencies, where signal penetration through common building materials is limited, and propagation is dominated by surface reflections and multipath effects \cite{vishwakarma2020micro}. Prior works in \cite{dardari2021nlos,buzzi2021radar,buzzi2022foundations,rihan2022spatial,mercuri2023reconfigurable,yasmeen2024around} have examined the use of RIS-assisted analog beamforming to enable programmable redirection of incident energy toward selected spatial regions beyond the radar antenna's field of view. Thus, RIS shifts the beamforming burden onto its low-power, low-cost passive-scattering surface compared to a reflectarray-based radar with a digital beamforming transceiver \cite{patyuchenko2013concept}.
Authors have explored the use of RIS to mitigate multipath effects in indoor localization and monitoring systems \cite{zhang2023multi,wymeersch2020radio,mercuri2023reconfigurable}. In \cite{aubry2021reconfigurable,buzzi2022foundations,zhang2023multi,chen2023robust}, the authors focused on idealized continuous‑phase RIS models, emphasizing far‑field scattering and simplified propagation assumptions without explicitly addressing the characterization of the RIS in the link budget analysis. This limitation was partially addressed in  \cite{liu2024equivalence}, where the authors used equivalence principles to model propagation in RIS-assisted environments. More accurate path‑loss models for RIS‑assisted wireless links were also investigated in \cite{tang2020wireless}. The authors of \cite{liu2021reconfigurable} demonstrated the potential of RIS to create non‑line‑of‑sight sensing paths, thereby enabling detection in shadowed regions. In \cite{wang2021received}, the authors showed RIS‑aided radar detection under various geometric configurations, showing significant gains in the signal-to-noise ratios (SNR). 

While these studies provide valuable theoretical insights, they often overlook hardware limitations such as unit‑cell non-idealities, quantization effects, power loss, and scattering behavior of real RIS. For example, the phase shifts introduced in RIS unit cells are typically coarsely quantized to reduce fabrication cost and complexity \cite{sahoo20231}. The 1-bit phase-quantized RIS, in particular, is the easiest and cheapest to build and hence is the most commonly studied in the literature \cite{jian2022reconfigurable, basar2024reconfigurable}. The discontinuous phase profiles from quantized RIS result in reduced aperture gain, and the formation of grating lobes \cite{khalil2025mitigating,singh2019controlling}. 
More recent studies by \cite{narayanan2024optimum} and \cite{khalil2025mitigating} proposed phase‑error mitigation and grating‑lobe suppression strategies for quantized RIS.

Despite these advances, experimental validation of RIS in radar scenarios remains limited.
Key scientific analysis of the enhancement of radar detection performance using a real RIS with phase-quantized unit cells is currently lacking in the literature. Specifically, when RIS is operated in radar scenarios, the signals must undergo two bistatic propagation paths: (1) the forward path from radar to RIS to target and (2) the backward path from target to RIS to radar. Hence, the scattering performance of the RIS, quantified by its bistatic radar cross-section (RCS) along the forward and backward paths, is a critical parameter governing the overall path loss. Prior work has simply assumed that RCS performance along both paths demonstrates angular symmetry. However, we show that the linear phase gradient on a finite-sized RIS designed to redirect an incident wave from a specific direction does not completely satisfy the constructive interference condition when the illumination direction is reversed. Consequently, the bistatic RCS peak values may differ for the forward and backward directions, leading to an angular asymmetry. This effect is further exacerbated by the quantization of unit cells and diffraction from small aperture sizes. The objective of this work is to analyze, quantitatively and qualitatively, the radar detection performance with a real quantized RIS. The novel contributions of our work, mapped against prior art, are presented in Table \ref{tab:literature}.

We summarize our contributions below: (1) Using aperture field theory, we derive closed‑form expressions for the scattered field from a two-dimensional RIS, explicitly modeling phase discontinuities arising from quantization. To bridge ideal aperture-field theory with practical RIS electromagnetic behavior, we incorporate equivalent surface currents extracted from full-wave simulations into the analytical RCS formulation. This hybrid analytical framework captures the impact of near-field interactions or mutual coupling, edge diffraction, and polarization conversion on the bistatic RCS. (2) We study the effect of 1-bit, 2-bit, and 3-bit phase quantization on RIS beam steering performance, with respect to the peak-to-specular ratio (PSLR), and the bistatic RCS across the forward (radar to RIS to target) and backward paths (target to RIS to radar). We highlight an interesting phenomenon: the bistatic RCS peak values of the RIS along the forward and backward paths are, in fact, \emph{not identical}, and must be individually characterized and factored into the overall path-loss model to carry out accurate link-budget analysis. (3) We evaluate the effect of practical RIS hardware impairments by modeling element-level amplitude and phase errors. Through Monte Carlo analysis, we quantify their impact on RCS mean-square error, SNR degradation, and beam-squint error, thereby clarifying how actual RIS responses affect radar sensing. (4) We present a fabricated $[16\times10]$ 1-bit RIS operating at 5.5 GHz and demonstrate electronically phase-controlled beam steering within an anechoic chamber. The measurements show strong agreement with theory and full-wave simulations. The RIS hardware enables us to experimentally study the impact of fabrication parameters such as quantization, lossy elements, and edge diffraction on the bistatic RCS of the RIS, beam-squint performance, and PSLR. (5) We experimentally validate RIS-assisted radar sensing in an anechoic chamber and real radio channels. The fabricated RIS enables electronically controlled beam steering toward non-specular regions, allowing recovery of micro-Doppler signatures from moving targets in non-line-of-sight (NLOS) conditions that are otherwise undetectable without RIS assistance. This is quantitatively verified using radar operative curves and repeatability analysis.

The remainder of the paper is organized as follows. We present the scattering field theory and derive closed-form bistatic RCS expressions for a phase-quantized RIS in Section~\ref{sec:theory} and study the numerical results in Section~\ref{sec:theoritical_results}. Section~\ref{sec:Full_Wave_Simulation} presents the full wave simulation results for RIS-enhanced radar detection of targets. Section~\ref{sec:measurement} describes the experimental setup and measurement results. Section~\ref{sec:real_env} presents the real environment enhanced target detection using RIS. Section~\ref{sec:conclusion} concludes the paper and outlines directions for future work.

\noindent\emph{Notation:} Scalars and vectors are represented by ordinary and bold font, respectively. $\hat{\mathbf{x}}$ denotes a unit vector with unit magnitude and direction along $x$ axis. $\theta$ is defined as the angle between a ray and the positive $z$ axis, while $\phi$ is the angle between the projection of a ray on the $xy$ plane and the positive $x$ axis. 
\begin{table*}[htbp]
\centering
\begin{threeparttable}
\small
\caption{ Comparison of prior work on RIS-enabled radar sensing and scattering with the proposed work. }

\begin{tabular}{p{1.2cm}|p{2cm}|p{2.1cm}|p{2.5cm}|p{1.5cm}|p{2.3cm}|p{2.2cm}}
\hline 
\textbf{Reference} & \textbf{Frequency} & \textbf{RIS Size} & \textbf{Phase Quantization} & \textbf{Minimum Switching Time} & \textbf{RCS Calculation of RIS} & \textbf{Hardware Validation} \\
\hline

\cite{buzzi2022foundations} & 3 GHz & 225 elements & continuous & NR & Analytical model & No \\
\hline

\cite{mercuri2023reconfigurable} & 24 GHz & generic array size & Discrete phase control (no quantization, pure geometric reflection) & NR & Radar equation including RIS element RCS & Metal reflectors instead of real RIS \\
\hline

\cite{aubry2021reconfigurable} & 10 GHz & 101×101, 133×133, 201×201 & continuous & NR & Array-based reflective gain (no RCS calculation) & No \\
\hline

\cite{chen2023robust} & 28 GHz & generic array size & 1-bit & NR & Bistatic channel-based RCS & No \\
\hline

\cite{rihan2022spatial} & Generic (wavelength $\lambda$, bandwidth $W$) & generic array size & Continuous phase \& amplitude & NR & Statistical target response model & No \\
\hline

\cite{xie2024ris} & Generic radar wavelength & generic array size & continuous & NR & No & No \\
\hline

\cite{vejling2025ris} & 30 GHz & 16×16 & continuous & NR & No & No \\
\hline

\cite{10878502} & 28 GHz & 16×16 & continuous & NR & No & No \\
\hline

\cite{yang2025frequency} & 30 GHz & 16×16 & continuous & NR & No & No \\
\hline
This Work & 5.5 GHz & 16×10 & 1-bit &100ns & Yes (using aperture field theory) & Yes \\
\hline
\end{tabular}
\label{tab:literature}
\begin{tablenotes}
\item[] Note: *NR = not reported
\end{tablenotes}
\end{threeparttable}
\end{table*}

\section{Theory}
\label{sec:theory}
\begin{figure}[htbp]
\centering
\includegraphics[scale=0.6]{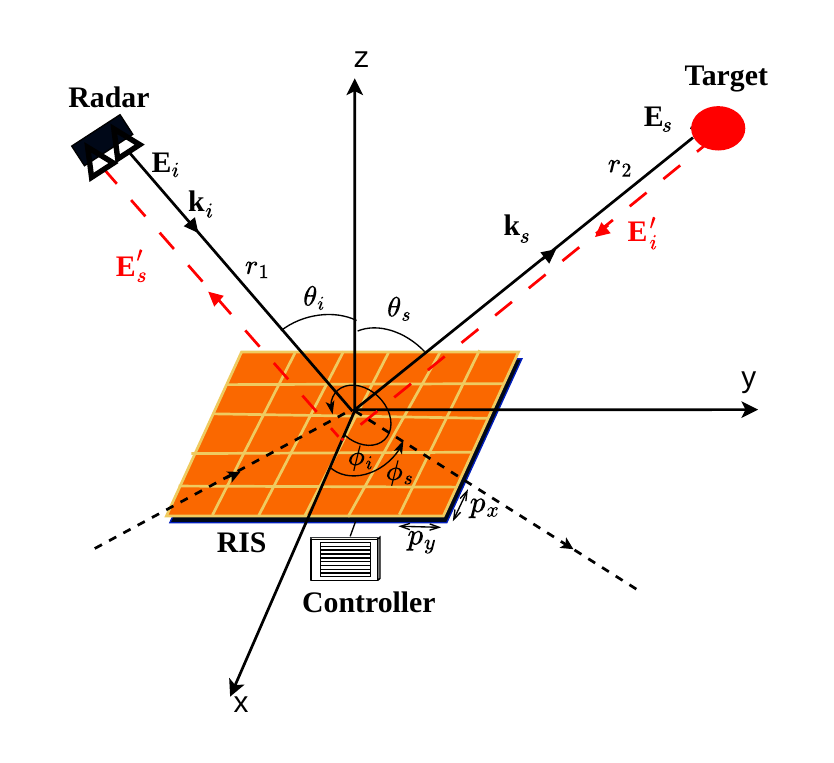}
\caption{RIS-enhanced radar system illustrating non-specular signal steering using RIS.}
\label{fig:illustration}
\end{figure}

This work aims to characterize the enhancement in radar detection performance achieved with a quantized RIS. We have considered a scenario where the radar transmits and receives through a beam directed at the RIS, as shown by the solid black arrow in Fig.~\ref{fig:illustration}. The RIS is configured to reflect the incident wavefront from the radar transmitter toward the target. The scattered echoes from the target are then redirected by the RIS back to the radar receiver, as shown by the dashed red arrow. Note that, in this scenario, the target is outside the mainlobe of the radar and hence there is no direct signal from the radar to the target and back.

We consider a two-dimensional (2D) RIS consisting of $M\times N$ uniformly spaced unit cells arranged in the $xy$-plane. Each unit cell has dimensions $p_x$ and $p_y$ along the $x$- and $y$-axes, respectively. We assume a transverse electric (TE) polarised electromagnetic plane wave, $\mathbf{E}_i$, with $E_0$ amplitude, incident upon the RIS at elevation and azimuth angles of $(\theta_i,\phi_i)$ as illustrated in Fig.~\ref{fig:illustration}. 
Here, $\mathbf{r}_{m,n}$ is the position vector of the center of the $(m,n)^{\text{th}}$ unit cell, $\eta_0$ is the wave impedance of free space, and the incident wave propagation vector is $\mathbf{k}_i = k_0(-\sin\theta_i \cos\phi_i \, \hat{\mathbf{x}} - \sin\theta_i \sin\phi_i \, \hat{\mathbf{y}} -\cos\theta_i \, \hat{\mathbf{z}})$ where $k_0$ is the free space wavenumber.  
The unit cells of an RIS are individually configured to direct the surface's main lobe towards the desired direction $(\theta_d,\phi_d)$  using $\Gamma_{m,n}$, the surface reflection coefficient at each $(m,n)^{th}$ unit cell and $\mathbf{k}_r$, the reflected wave vector.
According to Snell's law, $\mathbf{k}_r$ is given by $\mathbf{k}_r = k_0(\hat{\mathbf{k}}_i - 2(\hat{\mathbf{k}}_i \cdot \hat{\mathbf{z}})\hat{\mathbf{z}})$,
where $\hat{\mathbf{z}}$ is the surface normal vector of the RIS unit cell. Based on the field equivalence principles, the surface electric current density on the RIS element (at $z=0$) is
\begin{equation}
\begin{aligned}
    \mathbf{J}_{m,n} = \hat{\mathbf{z}} \times (\mathbf{H}_i + \mathbf{H}_r) = \\
    \hat{\mathbf{x}} \frac{E_0 }{\eta_0}\cos\theta_i(e^{-j \mathbf{k}_i \cdot \mathbf{r}_{m,n}}- \Gamma_{m,n}e^{-j \mathbf{k}_r \cdot \mathbf{r}_{m,n}}).
\end{aligned}
\end{equation}
To calculate the scattered electric field, we integrate the contributions of all unit cells over the RIS surface. In the far-field region, the scattered electric field from the RIS at any direction, $(\theta_s,\phi_s)$, due to surface electric currents is 
\begin{equation}
\begin{aligned}
E_{s_{\theta}} \simeq 
\frac{-j k_0 \eta_0\cos\theta_s \cos\phi_s p_x p_y   e^{-j k_0 r}}{4\pi r} 
\sum_{m=1}^{M} \sum_{n=1}^{N} \mathbf{J}_{m,n}
e^{j \mathbf{k}_s \cdot \mathbf{r}_{m,n}}
\end{aligned}
\label{eq:7}
\end{equation}

\begin{equation}
\begin{aligned}
E_{s_{\phi}} \simeq 
\frac{j k_0 \eta_0 p_xp_y\sin\phi_s e^{-j k_0 r}}{4\pi r} 
\sum_{m=1}^{M} \sum_{n=1}^{N} \mathbf{J}_{m,n}
e^{j \mathbf{k}_s \cdot \mathbf{r}_{m,n}}
\end{aligned}
\label{eq:8}
\end{equation}
where $\mathbf{k}_s$ is the scattered wave vector.
Similarly, surface magnetic current density, $\mathbf{M}_{m,n}$, can be computed from the tangential electric fields and will contribute towards the total scattered fields.
\subsection{Quantization of Phase}
If we assume an infinitely large, perfectly electric conductive planar surface, the reflection coefficient is given by $\Gamma_{m,n} = -1$ for all $m,n$. Instead, we consider an RIS where each element allows a programmable reflection coefficient, $\Gamma_{m,n} = A_{m,n} e^{j \Phi_{m,n}}$,
where $A_{m,n}$ and $\Phi_{m,n}$ are the electronically tunable amplitude and phase that modulate the contribution of each unit cell to the total scattered field. The unit cell phases required to direct the main lobe of the scattered field towards a desired $(\theta_d, \phi_d)$ can be calculated from the generalized Snell's law using
\begin{align}
\label{eq:snell}
    \Phi_{m,n} &= \mathbf{k}_i \cdot \mathbf{r}_{m,n} - \mathbf{k}_d \cdot \mathbf{r}_{m,n} \notag\\
    &= k_0 \Big[ (mp_{x} \sin\theta_i \cos\phi_i + np_{y} \sin\theta_i \sin\phi_i) \notag\\
    &\quad - (mp_{x}\sin\theta_d \cos\phi_d + np_{y}\sin\theta_d \sin\phi_d) \Big].
\end{align}
Substituting $\Gamma_{m,n}$ into the far-field approximation in \eqref{eq:7} and \eqref{eq:8}, the mainlobe of the scattered electric field can be directed towards the desired $(\theta_d,\phi_d)$ \cite{liu2024equivalence, balanis2024balanis}.
When observing the scattered field in the plane of incidence (the $yz$-plane, where $\phi = 90^\circ$), the spherical unit vector $\hat{\phi}$ is aligned with the $x$-axis, whereas $\hat{\theta}$ lies completely orthogonal to it. Consequently, the $E_{s_{\phi}}(\theta_s, \phi_s)$ and $E_{s_{\theta}}(\theta_s, \phi_s)$ components can be represented as co-polarized and cross-polarized scattered fields, respectively. The total magnitude of the scattered field is
\begin{equation}
    \label{eq:quantized_Esc}
    |\mathbf{E}_{s}(\theta_s, \phi_s)| = \sqrt{|E_{s_{\theta}}(\theta_s, \phi_s)|^2 + |E_{s_{\phi}}(\theta_s, \phi_s)|^2  }.
\end{equation}
The discussion above ignores the near-field interactions between the elements which are significant for small inter-element spacing and small sized apertures. To improve the accuracy of the proposed analysis, the equivalent surface current densities, $\textbf{J}_{m,n}$ and $\textbf{M}_{m,n}$, are extracted from full-wave simulated near-field E- and H-fields and used to evaluate both $E_{s_{\theta}}$ and $E_{s_{\phi}}$. Since the full-wave simulation inherently accounts for mutual coupling and other electromagnetic interactions among the RIS elements, these effects are incorporated into the proposed analytical framework.

The radiation pattern of the scattered fields from the RIS is governed by the amplitude and phase of each unit cell's reflection coefficient. We identify the following cases:\\
\noindent\textbf{Case 1: Continuous Amplitude and Phase Variation}
In this case, the main lobe of the scattered field can be directed along any direction within the RIS field of view, by tuning $A_{m,n}$ and $\phi_{m,n}$ to any values between $(0,1]$ and $[0,2\pi)$, respectively. \\
\noindent\textbf{Case 2: Constant Amplitude and Discrete Phase Variation} 
In hardware implementations, the amplitude and phase coefficients are quantized, i.e, can only be tuned to a fixed set of values due to fabrication constraints. Most commonly, the amplitudes are identical (e.g., $A_{m,n} = 1$) while the phase shifts are quantized to a finite set of levels, $2^b$, determined by $b$ number of control bits. 
More precise control of the main lobe requires lower quantization error and a higher $b$. However, implementing higher-bit quantization (e.g., 2-bit or 3-bit) in practical RIS hardware introduces several challenges, including increased circuit complexity, higher power consumption, and the need for precise phase control, which may involve multiple PIN diodes or varactor-based tuning networks per unit cell. Additionally, ensuring phase linearity and minimizing insertion loss across quantized states further increases design difficulty. Due to these limitations, 1-bit quantization remains a widely adopted choice in current RIS prototypes, offering a good trade-off between performance and hardware simplicity. Here, each unit cell can switch between two discrete phase states, with unit amplitude:
\begin{equation}
   \begin{aligned}
    \Phi_{m,n} \in \{0, \pi\} 
\quad \Rightarrow \quad 
\Gamma_{m,n} \in \{+1, -1\}.
\end{aligned} 
\end{equation}
A critical challenge in 1-bit quantized RIS is the occurrence of \emph{grating lobes} in the radiation pattern. To analyze this phenomenon, we consider the scattered field expression in \eqref{eq:quantized_Esc}. We determine the RIS unit cell phase coefficients to direct the main lobe along the desired direction \((\theta_d, \phi_d)\). Grating lobes occur at \((\theta_d^\ast, \phi_d^\ast)\) when the aperture gain at \((\theta_d^\ast, \phi_d^\ast)\) equals that at \((\theta_d, \phi_d)\) i.e.
\begin{align}
\label{eq:phi_tuning}
   \nonumber \max_{\{\Phi_{m,n}\}} \big| \mathbf{E}_s(\theta_d, \phi_d) \big|^2
= |\mathbf{E}_s(\theta_d^*, \phi_d^*) \big|^2 \\
\text{s.t.}  \Gamma_{m,n} \in \{+1, -1\}, \forall \, m,n .
\end{align}
It can happen if the phase difference across an effective period of the quantized RIS is an integer multiple of $2\pi$, i.e.,
\begin{subequations}
\begin{align}
   \Delta \Phi_{xp}= (k_{0}\sin\theta_{d} \cos\phi_{d} - k_{0}\sin\theta_i \cos\phi_i)P_{x} &= \pm 2\pi u \label{eq:16a}, \\
  \Delta \Phi_{yp} = (k_{0}\sin\theta_{d} \sin\phi_{d} - k_{0}\sin\theta_i \sin\phi_i)P_{y} &= \pm 2\pi v \label{eq:16b}.
\end{align}
\label{eq:grating_lobes_1}
\end{subequations}
Here, $\Delta \Phi_{xp}$ and $\Delta \Phi_{yp}$ denote the phase differences across the effective periodicities $P_x$ and $P_y$, respectively, and $u$ and $v$ are integer numbers. By substituting ($\theta_{d}, \phi_{d}$) with ($\theta_{d}^*, \phi_{d}^*$), the wavenumber expression ($k = 2\pi/\lambda$) and then rearranging, the beamforming conditions become:
\begin{subequations}
\begin{align}
    \sin\theta_{d}^* \cos\phi_{d}^* - \sin\theta_i \cos\phi_i &= \frac{\pm u\lambda}{P_x} \label{eq:17a} \\
    \sin\theta_{d}^* \sin\phi_{d}^* - \sin\theta_i \sin\phi_i &= \frac{\pm v\lambda}{P_y} \label{eq:17b}
\end{align}
\label{eq:grating_lobes}
\end{subequations}
 Without quantization, the effective periodicities of the RIS along the two axes are the element spacings $p_x$ and $p_y$, which are less than $\lambda/2$, and hence no grating lobes appear. However, with 1-bit quantization, the quantized phase distribution results in $P_x >p_x$ and $P_y>p_y$ along some stretches of the aperture. As a result, 
 the periodicity exceeds $\lambda/2$ leading to grating lobes in the radiation pattern. 
\subsection{Radar Cross Section of RIS }
In RIS-assisted radar systems, signal propagation occurs in two stages: the \textit{forward path} (radar transmitter to RIS to target along the solid black arrow in the figure) and the \textit{backward path} (target to RIS to radar receiver along the dashed red arrow in the figure). Here, $r_1$ and $r_2$ denote the distances from the RIS to the radar and the target, respectively. The unit cells of the passive RIS can be tuned to only one specific combination of $A_{m,n}$ and $\phi_{m,n}$ at a time. Hence, the same tuning configuration governs both the forward and backward paths. Consequently, the design of the RIS phase profile must account for its dual impact on both signal transmission and echo reception, as illustrated in Fig.~\ref{fig:illustration}.

The RCS along a specified direction $(\theta_s,\phi_s)$ is defined as the ratio of the scattered power along that direction to the incident power along the illumination direction $(\theta_i,\phi_i)$. For the \textit{forward path}, the RCS of RIS is expressed as
\begin{equation}
    \sigma_f(\theta_s,\phi_s;\theta_i,\phi_i) = 4\pi r_2^2 \frac{\left| \mathbf{E}_s(\theta_s,\phi_s) \right|^2}{\left| \mathbf{E}_i(\theta_i,\phi_i) \right|^2},
    \label{eqn:RCS_forward}
\end{equation}
where $\mathbf{E}_s(\theta_s,\phi_s)$ and $\mathbf{E}_i(\theta_i,\phi_i)$ denote the magnitudes of the scattered and incident electric fields, respectively. The RCS for the \textit{backward path} is defined as the ratio of the scattered power along the incident direction due to the incident field along the previous scattered direction, given by
\begin{equation}
    \sigma_b(\theta_i,\phi_i;\theta_s,\phi_s) = 4\pi r_1^2 \frac{\left| \mathbf{E}'_s(\theta_i,\phi_i) \right|^2}{\left| \mathbf{E}'_i(\theta_s,\phi_s) \right|^2}.
    \label{eqn:RCS_backward}
\end{equation}
Here, the incident field, $\mathbf{E}'_i(\theta_s,\phi_s)$,  emanates from the target located at $(\theta_s,\phi_s)$ and falls upon the RIS while the scattered field, $\mathbf{E}'_s(\theta_i,\phi_i)$, is the resultant field from the RIS falling upon the radar located at $(\theta_i,\phi_i)$. 

It is worth noting that for ordinary scatterers operating in linear time invariant radar systems, the bistatic RCS satisfies electromagnetic reciprocity - meaning that the bistatic scattering amplitude is symmetric with respect to the interchange of incident and observation directions, leading to
\begin{equation}
\sigma_f(\theta_s,\phi_s;\theta_i,\phi_i) = \sigma_b(\theta_i,\phi_i;\theta_s,\phi_s),
\end{equation}
for identical polarization states. However, for RIS, the phase condition in \eqref{eq:snell} may not be satisfied when $\theta_i,\phi_i$ are interchanged with $\theta_d,\phi_d$. Despite that, for radar scenarios, the phase condition calculated for the forward path is kept fixed for both the forward and backward paths. The condition is satisfied only in specific scenarios in which both sets of angles are exactly equal. Further, this angular asymmetry is compounded under quantization and for smaller aperture sizes, which are affected by diffraction. 

Consider a monostatic radar with transmit power, $P_{tx}$, radar antenna gain, $G_a$, and mean noise floor, $\mathcal{N}$, used to detect a target of RCS, $\sigma_t$, that is not directly within the radar's field of view. A finite-aperture RIS is programmed to direct its main lobe towards the target to enhance the detection. 
Based on the above discussion, the SNR at the radar receiver with RIS enhancement can be evaluated using the expression given by  
\begin{align}
\label{eq:snr}
\text{SNR} = \frac{P_{tx}G_a^2(\theta_i.\phi_i)\sigma_f(\theta_s,\phi_s;\theta_i,\phi_i)\sigma_t\sigma_b(\theta_i,\phi_i;\theta_s,\phi_s)\lambda^2}{(4\pi)^5r_1^4r_2^4\mathcal{N}},
\end{align}
based on Frii's bistatic radar range equation \cite{willis2005bistatic}.
Unlike the two-way propagation path encountered by conventional radar, the incorporation of the RIS introduces a path loss that accounts for both $\sigma_f$ and $\sigma_b$, and four-way propagation from radar to RIS, to the target, back to RIS, and then to radar. 

\section{Theoretical Results}
\label{sec:theoritical_results}
This section presents the RCS of RIS, focusing on the effects of phase quantization and RIS aperture size.

\subsection{Bistatic RCS of RIS Along Forward and Backward Paths}
\begin{figure*}[htbp]
\centering
\includegraphics[width=490mm,height=70mm,keepaspectratio]{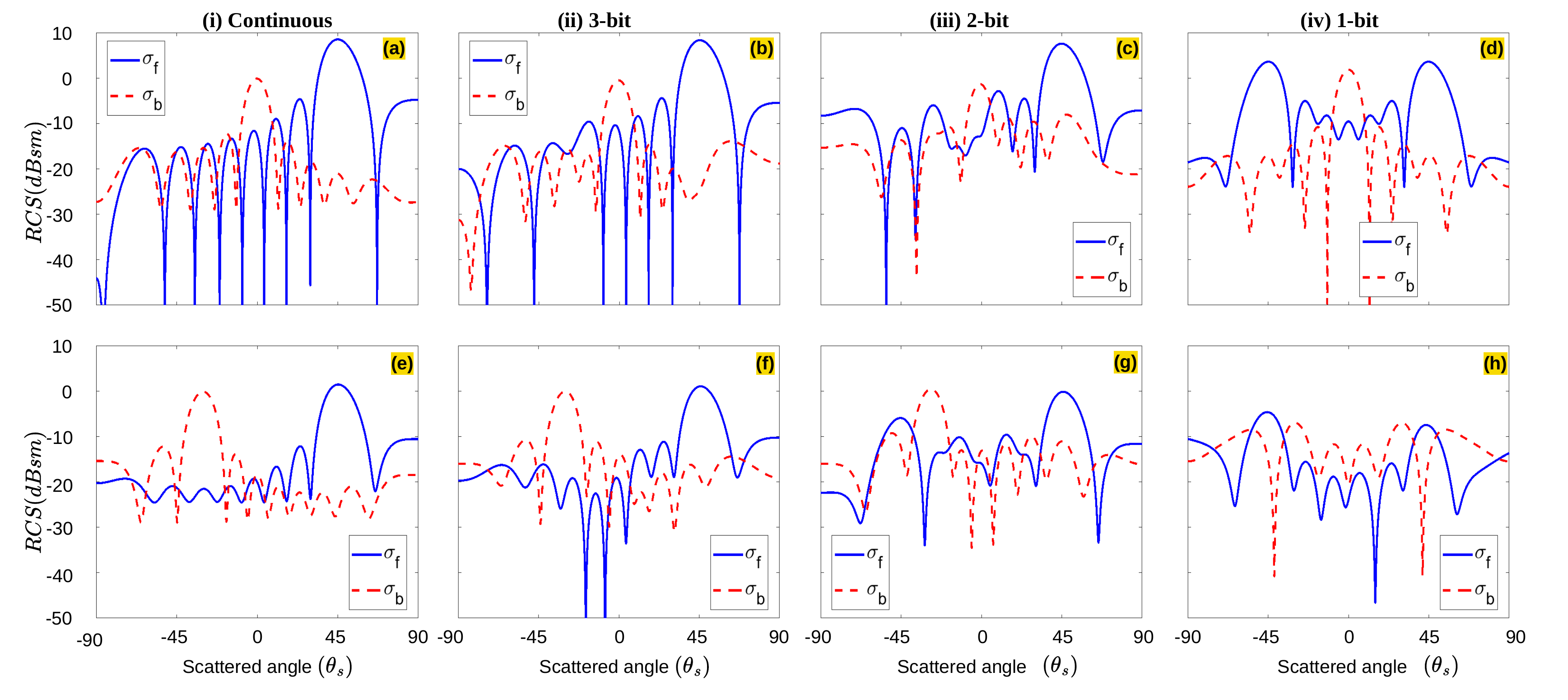}
\caption{RCS (dBsm) of the RIS in the $\phi_s = 90^\circ$ plane for forward and backward scattering. The columns represent phase quantization levels: (i) continuous, (ii) 3-bit, (iii) 2-bit, and (iv) 1-bit. The first row (a-d) corresponds to ($\theta_i = 0^\circ,\theta_d = 45^\circ$) while second row (e-h) corresponds to ($\theta_i = -30^\circ,\theta_d = 45^\circ$).} 
\label{fig:N_bit_discretized}
\end{figure*}

We consider an RIS consisting of $[16\times10]$ unit cells uniformly at $0.016$m, and operating at 5.5 GHz.  
The RIS is assumed to be in the far field of both the radar and the target and placed on the $xy$ plane. The RCS patterns are computed for the $yz$ plane for varying $\theta_s$ while $\phi_i=\phi_s = 90^\circ$ based on the aperture field theory discussed in the previous section. We derive the RCS patterns for forward and backward paths under two scenarios. In the first scenario,  $\theta_i = 0^{\circ}$ with results shown in the top row of Fig.\ref{fig:N_bit_discretized} and in the second scenario, $\theta_i = -30^{\circ}$, with results shown in the the bottom row of Fig.\ref{fig:N_bit_discretized}.
For both scenarios, the RIS elements are tuned for the main lobe to be directed along $\theta_d=45^{\circ}$. 
The results are presented for RIS with unit cells capable of continuous phase shifts and for 1-bit, 2-bit, and 3-bit quantizations. 
Fig.\ref{fig:N_bit_discretized}a shows that for $\theta_i=0^{\circ}$, the main lobe of the forward RCS pattern peaks at the desired angle $45^\circ$, with $\sigma_f$= 8.5 dBsm with a PSLR of 21.2 dB. In the backward scattering case, $\sigma_b$ for $\theta_i = 45^{\circ},\theta_d=0^{\circ}$ is -0.1 dBsm, which is much lower than  $\sigma_f$ for $\theta_i =0^{\circ},\theta_d=45^{\circ}$ due to weak constructive interference and readjustment in the sidelobe pattern. 

Next, we consider the effect of discrete phase quantization on RIS beamforming in Fig.\ref{fig:N_bit_discretized}b-d. The results show that the $\sigma_f$ pattern of the $b = 3$-bit-quantized RIS in Fig.\ref{fig:N_bit_discretized}b most closely resembles the continuous case in Fig.\ref{fig:N_bit_discretized}a. However, for lower $b$, the quantization error increases, leading to grating lobes. This is most prominent for the 1-bit quantized RIS in Fig.\ref{fig:N_bit_discretized}d where the main lobe peaks at $45^{\circ}$, but a grating lobe appears at $-45^{\circ}$ (consistent with the dual beam pattern in \cite{yasmeen2024around}). 

In each of these quantization cases, the main lobe of the backward RCS peaks at $0^{\circ}$. However, the peak $\sigma_b$ never matches the peak $\sigma_f$. Similarly, there is a decrease in the PSLR for the lower $bit$-quantization. Next, we consider the second scenario where the forward wavefront is incident on the RIS $\theta_i = -30^\circ$ as shown in the bottom row of Fig.\ref{fig:N_bit_discretized}. Again, the unit cells are tuned to steer the reflected main lobe to $\theta_d=45^{\circ}$. 
For the continuous phase control of the unit cells, the $\sigma_f$ pattern again peaks at the desired angle of $45^{\circ}$ in Fig.\ref{fig:N_bit_discretized}e. 
However, the $\sigma_f$ for $\theta_i = -30^{\circ},\theta_d = 45^{\circ}$ is 1.4 dBsm which is significantly lower than the $\sigma_f$ for $\theta_i = 0^{\circ},\theta_d = 45^{\circ}$ which was shown in Fig.\ref{fig:N_bit_discretized}a. 
Next, we present the $\sigma_f$ and $\sigma_b$ patterns for 3-bit, 2-bit, and 1-bit quantized units tuned for the same $\theta_i = -30^{\circ}$ and $\theta_d = 45^{\circ}$ in Fig.\ref{fig:N_bit_discretized}f-h. The results again show that the 3-bit quantized RIS demonstrates the closest similarity with the unquantized RIS in Fig.\ref{fig:N_bit_discretized}f, while the 1-bit  RIS shows a significantly lower $\sigma_f$ peak than the remaining three cases. 

Table~\ref{table:simulated_RCS} presents the peaks of $\sigma_f$ and $\sigma_b$ patterns for unit cells tuned for different combinations of $\theta_i$ and $\theta_d$ and for different phase quantization levels. The results show that $\sigma_f = \sigma_b$ only when the $\theta_i =\theta_d = 0^{\circ}$ (first row of the table). This is because of the angular symmetry in the phase-gradient equation in \eqref{eq:snell}, which results in mirror-like RCS patterns.  
\begin{table}[htbp]
\centering
\caption{Bistatic RCS \{$\sigma_f$, $\sigma_b$\} for different quantizations for $[16\times10]$ RIS.}
\setlength{\tabcolsep}{3pt}
\renewcommand{\arraystretch}{1.05}

\resizebox{\columnwidth}{!}{
\begin{tabular}{cc|cc|cc|cc|cc|cc}
\hline
$\theta_i$ & $\theta_d$
& \multicolumn{2}{c}{Cont.}
& \multicolumn{2}{c}{3-bit}
& \multicolumn{2}{c}{2-bit}
& \multicolumn{2}{c}{1-bit(ideal)} 
& \multicolumn{2}{c}{1-bit(actual)} \\
\cline{3-12}
& 
& $\sigma_f$ & $\sigma_b$
& $\sigma_f$ & $\sigma_b$
& $\sigma_f$ & $\sigma_b$
& $\sigma_f$ & $\sigma_b$
&$\sigma_f$ & $\sigma_b$ \\
\hline\hline

$0^\circ$ & $0^\circ$
& 8.5 & 8.5 & 8.5 & 8.5 & 8.5 & 8.5 & 8.5 & 8.5 & 7.4 & 7.4  \\

$0^\circ$ & $15^\circ$
& 8.5 & 2.9 & 8.2 & 2.7 & 7.5 & 2.3 & 4.9 & 4.7& 3.5 & 3.7  \\

$0^\circ$ & $30^\circ$
& 8.5 & 1.4 & 8.2 & 1.4 & 7.6 & 0.9 & 4.0 & 3.3& 3.2 & 3.8  \\

$0^\circ$ & $45^\circ$
& 8.5 & -0.1 & 8.3 & -0.5 & 7.5 & -1.4 & 3.5 & 1.8& 1.5  & 2.7  \\
$-30^\circ$ & $0^\circ$
& 1.4 & 8.5 & 1.4 & 8.2 & 1.2 & 6.0 & 4.0 & 4.6 & 3.6 & 3.6 \\

$-30^\circ$ & $15^\circ$
& 1.4 & 2.9 & 1.1 & 2.6 & 0.9 & 1.1 & -1.0 & -1.1& 3.1 & 2.6  \\

$-30^\circ$ & $30^\circ$
& 1.4 & 1.4 & 1.4 & 1.4 & 0.9 & 1.2 & -0.8 & -0.8& 6.1  &  2.1 \\

$-30^\circ$ & $45^\circ$
& 1.4 & -0.1 & 0.9 & -0.1 & -0.2 & 0.1 & -7.6 & -6.9 & 3.1  & -3.0  \\
\hline
\end{tabular}}
 \label{table:simulated_RCS}
\end{table}
Interestingly, in this case, quantization does not degrade the RIS's performance. When we consider phase continuity scenarios (first column), the $\sigma_f$ remains fixed for specific $\theta_i$ regardless of the choice of $\theta_d$. 

\begin{figure}[htbp]
\centering
\includegraphics[width=90mm,height=85mm,keepaspectratio]{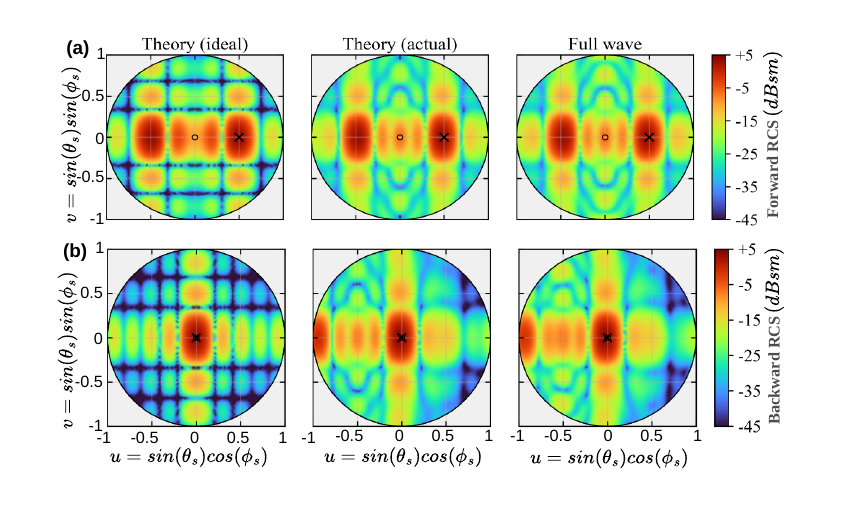}
\caption{Comparison of 2D-RCS of 1-bit RIS (a) forward and (b) backward path for (i) theory(ideal) (ii) theory(actual) and (iii) full-wave simulation for incident angle $\theta_i=0$ and desired angle $\theta_d=30^\circ$.}
\label{fig:2D_RCS_comparison}
\end{figure}

However, when we introduce quantization to the unit cells, we observe that the $\sigma_f$ varies with $\theta_d$ even for a fixed $\theta_i$. Even more significant variation in the $\sigma_f$ is observed as we lower the quantization levels, $b$, with the greatest degradation in RCS for 1-bit RIS and the least degradation for 3-bit RIS because of the phase errors introduced in \eqref{eq:snell}. Furthermore, for the 1-bit case, two scenarios are considered: the ideal 1-bit RCS obtained from theory, and the actual 1-bit RCS obtained from the hybridized analytical framework. The results presented in Fig.\ref{fig:2D_RCS_comparison}-Fig.\ref{fig:rev_Pol} show that the actual 1-bit RCS performance degrades compared with the ideal case due to near-field electromagnetic effects. The results also clearly demonstrate that $\sigma_f$ is not equal to $\sigma_b$ for most situations because of the RIS's finite aperture size and the complex interference patterns that result from reversing the directions of illumination and reflection. The coarse quantization of the unit cells results in further mismatch in the RCS values as it can be observed that the $\{\sigma_f,\sigma_b\}$ for $(\theta_i = 0^{\circ};\theta_d=30^{\circ})$ is equal to $\{\sigma_b,\sigma_f\}$ for $(\theta_i = -30^{\circ};\theta_d=0^{\circ})$ for unquantized RIS and 3-bit quantized RIS. For lower-bit quantization, constructive interference is weaker in the backward direction, resulting in lower peak RCS values and modified sidelobe patterns. Although idealized analytical models predict differing values for $\sigma_f$ and $\sigma_b$ due to the finite-bit
implementation and omission of mutual coupling and structural scattering, the physical macroscopic reciprocity must hold for a passive RIS. Using a modulo-$360^\circ$ phase-wrapping, we observe that the 1-bit phase code optimized specifically to steer a beam from $\theta_i \rightarrow \theta_d$ is mathematically distinct from the code optimized to steer for the backward path. Because the radar scenario forces the RIS to remain frozen in the forward-optimized state, the round-trip echo is evaluated against a sub-optimal backward code, dictating the peak round-trip gain, even while the localized bistatic intersection remains perfectly reciprocal.
To verify the angular redistribution caused by 1-bit quantization, Fig.~\ref{fig:2D_RCS_comparison} presents the 2D RCS maps over the visible angular region for the forward and reverse paths for $\theta_i=0^\circ$ and $\theta_d=30^\circ$. The maps are plotted using $u=\sin\theta_s\cos\phi_s$ and $v=\sin\theta_s\sin\phi_s$, thereby covering the complete range of $\phi_s$. The top row shows a 2D plot of the RCS for the forward path, while the bottom row shows it for the backward path. The columns present the ideal 1-bit aperture-field model, the actual-current theory, and the full-wave result. The results show that 1-bit phase coding introduces additional quantization lobes, while the main-lobe direction and sidelobe structure predicted by the analytical model are preserved. Compared with the ideal analytical results, the hybrid analytical results and full-wave simulations exhibit bigger specular lobes and variations in sidelobe levels due to mutual coupling, finite-aperture effects, and nonuniform current distributions. The close agreement between the theory (actual) and full-wave simulation confirms that the proposed model captures the dominant forward and backward RCS behavior.
\begin{table}[htbp]
\centering
\caption{Peak-to-Specular Lobe Ratio (PSLR (dB)) for different RIS quantizations along forward (F) and backward (B) paths for $(\theta_i=0^\circ;\theta_d)$.}
\setlength{\tabcolsep}{3.2pt}
\renewcommand{\arraystretch}{1.05}

\resizebox{\columnwidth}{!}{
\begin{tabular}{c|cc|cc|cc|cc|cc}
\hline
\multirow{2}{*}{$\theta_d$}
& \multicolumn{2}{c|}{Continuous}
& \multicolumn{2}{c|}{3-bit}
& \multicolumn{2}{c|}{2-bit}
& \multicolumn{2}{c|}{1-bit(ideal)}
& \multicolumn{2}{c}{1-bit(actual)} \\
\cline{2-11}
& F & B
& F & B
& F & B
& F & B
& F & B \\
\hline\hline

$15^\circ$ & 15.6 & 11.2 & 15.1 & 10.9 & 20.2 & 7.1 & 8.6 & 13.5 & 4.8 & 5.1 \\
$30^\circ$ & 18.1 & 14.1 & 18.5 & 10.8 & 13.2 & 4.7 & 14.2 & 16.1 & 9.7 & 11 \\
$45^\circ$ & 21.2 & 16.1 & 20.2 & 15.8 & 20.1 & 12.3 & 14.5 & 18.9 & 0 & 2.6 \\

\hline
\end{tabular}}
\label{table:pslr_sim}
\end{table}
Table~\ref{table:pslr_sim} quantifies the PSLR for different RIS quantizations. Here, the peak RCS is obtained at $\theta_s = \theta_d$, while the RCS at the specular angle is obtained from $\theta_s = \theta_i$. Hence, the PSLR is distinct from the peak-to-sidelobe ratio metric used in conventional array processing. In the continuous case, each element provides an ideal phase response, enabling precise wavefront shaping and constructive interference in the desired direction while effectively suppressing sidelobes. This results in consistently high PSLR values in both forward and backward scattering. When phase quantization is introduced, discrete phase states result in phase errors relative to the ideal continuous distribution as discussed in Section \ref{sec:theory}A. The signal from the main lobe is redistributed into sidelobes, thereby degrading the PSLR. 
For the 1-bit RIS, the forward PSLR is significantly lower than in the continuous case. The results highlight a fundamental trade-off between hardware complexity and RIS performance. While low-bit RIS designs reduce implementation cost and power consumption, they introduce significant phase quantization errors that limit specular suppression capability. The 1-bit RIS produces grating lobes, resulting in a lower PSLR. In the backward direction, since the illumination angle is not normal to the RIS plane, grating lobes are not produced, resulting in a relatively higher PSLR.
\subsection{Co- and Cross-Polarized Scattered Fields}
\begin{figure}[htbp]
\centering
\includegraphics[width=90mm,height=60mm]{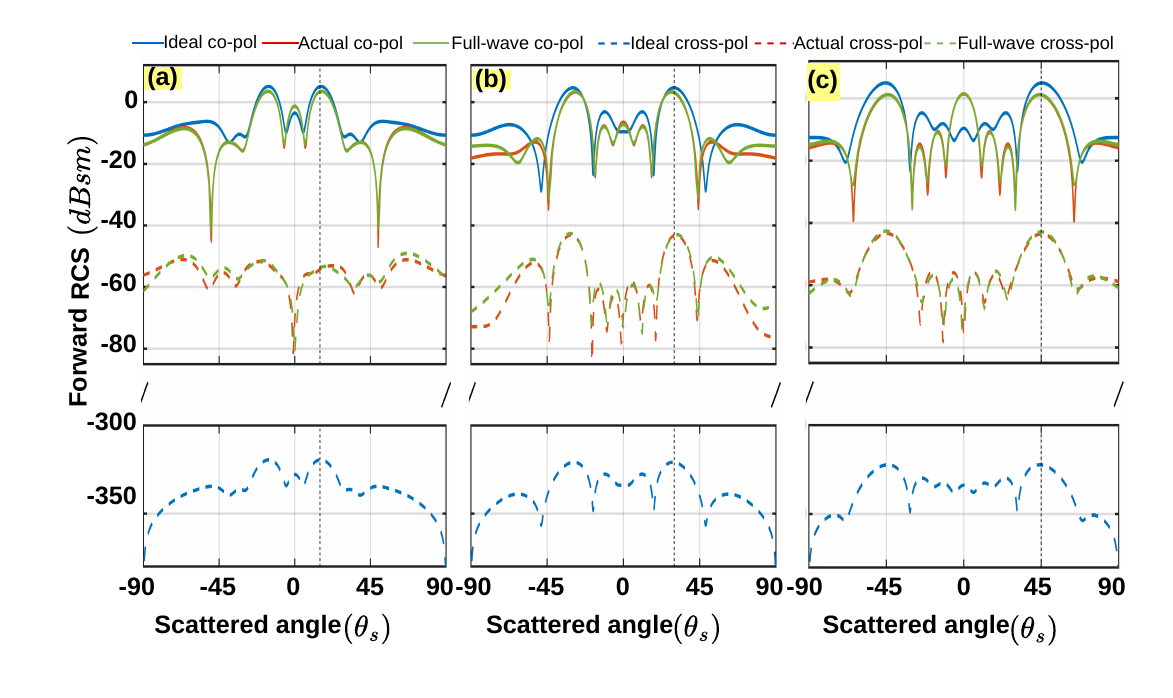}
\caption{Comparison of forward RCS of 1-bit RIS polarization from theory (ideal) and theory (actual) and full-wave simulation for incident angle $\theta_i=0$ and $\theta_d $ (a) $15^\circ$, (b) $30^\circ$, and (c) $45^\circ$.}
\label{fig:fwd_Pol}
\end{figure}

\begin{figure}[htbp]
\centering
\includegraphics[width=90mm,height=130mm,keepaspectratio]{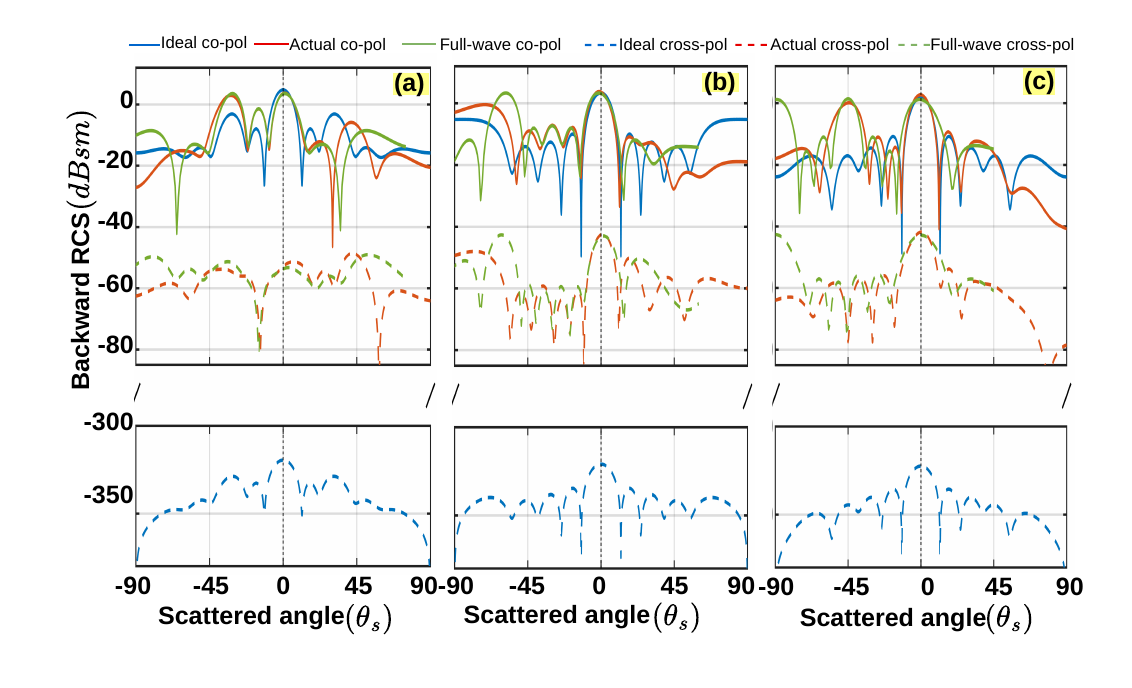}
\caption{Comparison of backward RCS of 1-bit RIS polarization from theory (ideal)and theory (actual) and full-wave simulation for incident angle $\theta_i=0$ and $\theta_d $ (a) $15^\circ$, (b) $30^\circ$, and (c) $45^\circ$.}
\label{fig:rev_Pol}
\end{figure}

To evaluate the polarization purity of the scattered field, the co-polarized ($E_{s_{\phi}}(\theta_s, \phi_s) $) and cross-polarized ($E_{s_{\theta}}(\theta_s, \phi_s) $) components of the forward and reverse RCS are evaluated and compared in Figures~\ref{fig:fwd_Pol} and~\ref{fig:rev_Pol}. 

 The results are shown for three desired reflection angles, $\theta_d=15^\circ$, $30^\circ$, and $45^\circ$, with normal incidence $\theta_i=0^\circ$. For both the forward and backward cases, the co-polarized RCS exhibits a dominant beam in the desired direction $\theta_d$. The cross-polarized RCS remains much lower than the co-polarized component, showing that the surface mainly preserves the incident polarization and produces only weak polarization conversion. The difference between the theory(ideal) and theory(actual) RCS is attributed to the practical current distribution, which includes the non-ideal effects of aperture.

\subsection{Effect of Aperture Size of RIS}
Next, we consider the effect of RIS aperture size on $\sigma_f$ and $\sigma_b$. We again consider an RIS aligned along the $xy$ plane at 5.5 GHz and compute the RCS from the patterns generated for the $\phi_s = 90^\circ$ plane. Table~\ref{table:RIS_of_diff_sizes} presents the peak $\sigma_f$ and $\sigma_b$ for different combinations of $(\theta_i;\theta_d)$ for a 1-bit RIS.
For $\theta_i = 0^\circ$ and $\theta_d = 0^\circ$, $\sigma_f=\sigma_b$ for all aperture sizes, indicating symmetric scattering. Further, $\sigma_f$ and $\sigma_b$ increase significantly with aperture size, consistent with array theory in \eqref{eq:7} and \eqref{eq:8}. As noted previously, both $\sigma_f$ and $\sigma_b$ vary when $\theta_d$ is changed for a fixed aperture size. However, an RIS with a large aperture (for e.g. the $[32\times10]$ RIS) still has high $\sigma_f$ and $\sigma_b$ even for unfavorable combinations of $\theta_i$ and $\theta_d$. Under oblique incidence ($\theta_i = -30^\circ$), the asymmetry between forward and backward scattering becomes more pronounced.  Overall, a smaller RIS aperture yields lower RCS values for both $\sigma_f$ and $\sigma_b$, especially at high $\theta_d$.  
\begin{table}[htbp]
\centering
\caption{Bistatic 1-bit RCS \{$\sigma_f$, $\sigma_b$\} for different RIS sizes.}
\setlength{\tabcolsep}{7pt} 

\resizebox{\columnwidth}{!}{%
\begin{tabular}{cc|cc|cc|cc}
\hline
$\theta_i$ & $\theta_d$
& \multicolumn{2}{c|}{$[8\times10]$}
& \multicolumn{2}{c|}{$[16\times10]$}
& \multicolumn{2}{c}{$[32\times10]$} \\
\cline{3-8}
& 
& $\sigma_f$ & $\sigma_b$
& $\sigma_f$ & $\sigma_b$
& $\sigma_f$ & $\sigma_b$ \\
\hline\hline

$0^\circ$ & $0^\circ$ 
& 2.4 & 2.4 & 8.5 & 8.5 & 14.5 & 14.5 \\
$0^\circ$ & $15^\circ$ 
& -0.5 & -0.9 & 4.9 & 4.7 & 10.5 & 10.1  \\
$0^\circ$ & $30^\circ$ 
& -0.7 & -2.0 & 4.0 & 3.3 & 10.7 & 9.5 \\
$0^\circ$ & $45^\circ$ 
& -0.8 & -4.1 & 3.5 & 1.8 & 10.7 & 7.6 \\
$-30^\circ$ & $0^\circ$ 
& -2.0 & -0.7 & 4.0 & 4.6 & 9.5 & 10.7 \\
$-30^\circ$ & $15^\circ$ 
& -5.0 & -4.0 & -1.0 & -1.1 & 3.8 & 4.7  \\
$-30^\circ$ & $30^\circ$
& -5.4 & -5.4 & -0.8 & -0.8 & 4.2 & 4.2 \\
$-30^\circ$ & $45^\circ$
& -8.6 & -10.4 & -7.6 & -6.9 & 0.8 & -0.8 \\
\hline
\end{tabular}
}
\label{table:RIS_of_diff_sizes}
\end{table}
\subsection{SNR Performance with Four-Way RIS Propagation}
We consider a monostatic radar operating at 5.5 GHz with transmit power $P_{tx}=1$ mW and antenna gain $G_a=12$ dBi. These parameters are chosen to be consistent with the experimental setup discussed in the following section. The target radar cross section is fixed at $\sigma_t=1$ dBsm. Receiver noise is modeled as $N_0=-105$ dBm (based on the experimental apparatus used in the following section). The RIS is located between the radar and target with distances $r_1=3$ m and $r_2=3$ m, respectively, as shown in the figure. The RIS unit cells are tuned to steer its mainlobe toward the target direction, $\theta_d$, for incidence angle $\theta_i=0^\circ$, and its scattering behavior is characterized by forward and backward RCS $\sigma_f(\theta_d;\theta_i)$ and $\sigma_b(\theta_i;\theta_d)$ as discussed in Table \ref{table:RIS_of_diff_sizes}.
\begin{figure}[htbp]
\centering
\includegraphics[scale =0.2]{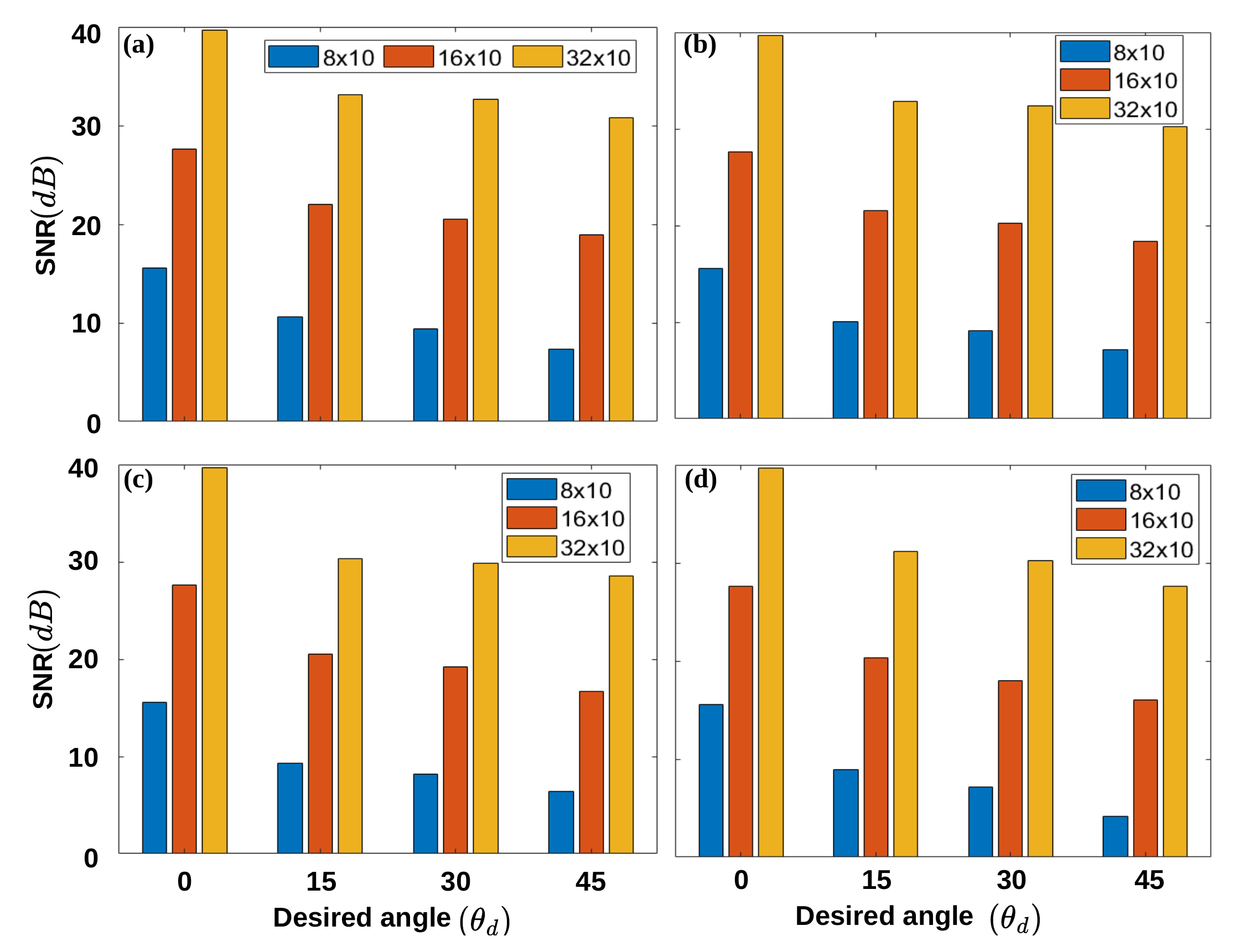}
\caption{SNR performance for RIS with (a) continuous, (b) 3-bit, (c) 2-bit, and (d) 1-bit phase quantization at $\theta_i = 0^\circ$ for varying RIS sizes.}
\label{fig:SNR_bar_graph}
\end{figure} 
We calculate the SNR for the RIS-enhanced radar scenarios using \eqref{eq:snr}, and the results are presented in Fig.~\ref{fig:SNR_bar_graph} for different aperture sizes and phase quantization levels of the unit cells. As anticipated, the SNR is highest for the 32-element RIS, followed by the 16- and 8-element RISs, across all quantization cases. This behavior is attributed to the larger effective aperture, which enhances the reflected signal power.
Additionally, the SNR declines as $\theta_d$ increases, since steering the beam toward larger angles reduces the effective $\sigma$ in the target direction. We also observe from Fig.~\ref{fig:SNR_bar_graph}a and Fig.~\ref{fig:SNR_bar_graph}b that the 3-bit is very close to continuous SNR values. However, coarse phase resolution introduces grating lobes and degrades the RCS pattern, leading to a significant decrease in SNR.

\subsection{Effect of Phase and Amplitude Errors}
In the absence of impairments, $\phi_{mn}^{\mathrm{ideal}}$ is the ideal quantized RIS phase, and the amplitude is one for each $(m,n)^{th}$ element.
We study the hardware impairments on SNR due to amplitude and phase errors at the element level by modeling the  reflection coefficient as
\begin{equation}
\Gamma_{mn}=
A_{mn}\exp\left[j\left(\phi_{mn}^{\mathrm{ideal}}+\epsilon_{mn}\right)\right].
\end{equation}
Here, $A_{mn}$ is the random amplitude coefficient modeled as a positive unit-mean random coefficient with variance, $\sigma_A^2$. The phase error, $\epsilon_{mn}$, is modeled as a uniformly distributed bounded error, where $\Delta\phi_{\max}$ is the maximum phase-error bound. It is swept from $0^\circ$ to $20^\circ$ for the statistical experiments. We consider two scenarios. First, we change the amplitude variance while keeping the phase error zero. Then we set $A_{mn}=1$ and change the phase error.
For each scenario, we perform Monte Carlo simulations to calculate the forward and reverse RCS and compute the SNR using (13). 
\begin{figure}[htbp]
\centering
\includegraphics[scale=0.25]{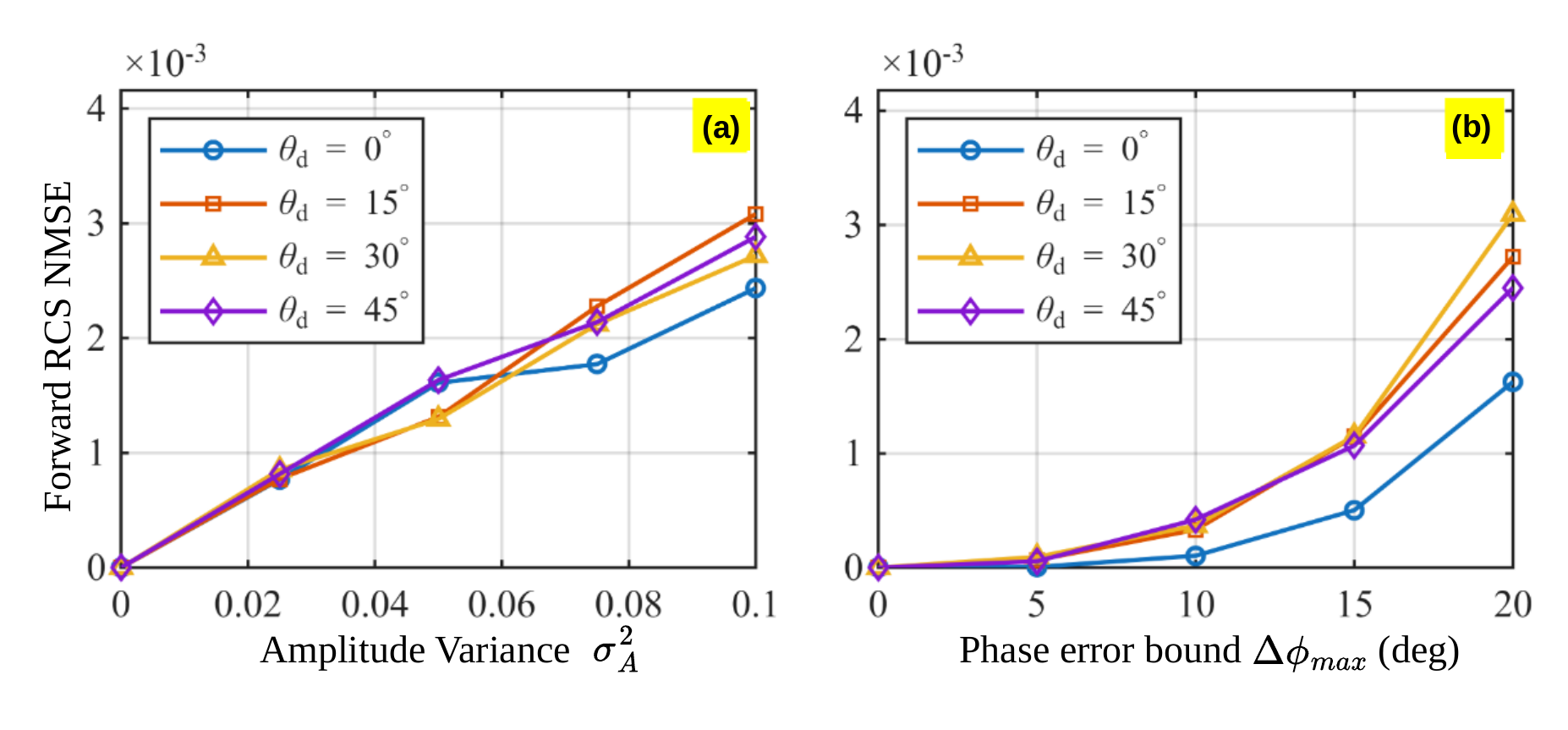}
\caption{Forward RCS error of the $[16\times10]$ 1-bit RIS for $\theta_i=0^\circ$ and different desired angles $\theta_d$, (a) show the RCS NMSE versus amplitude variance $\sigma_A^2$, (b) RCS NMSE versus phase-error $\Delta\phi$.}
\label{fig:rcs_amp_phase_variation}
\end{figure}
The normalized mean square error of RCS presented in Fig.\ref{fig:rcs_amp_phase_variation} shows that both amplitude and phase errors degrade the RIS signal strength. 
As $\Delta\phi_{\max}$ increases, the phase alignment across the RIS aperture becomes less accurate which reduces coherent combining in the desired direction and increases the RCS error. 
\begin{figure}[htbp]
\centering
\includegraphics[scale=0.35]{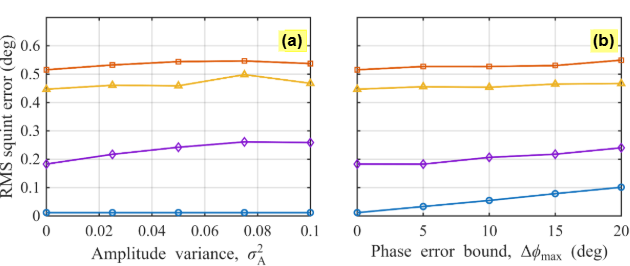}
\caption{Beam-squint error of 16$\times$10 RIS for 1-bit phase quantization with $\theta_i$=0 and desired angle $\theta_d$ for (a) fixed  $\Delta \phi=0^\circ$ amplitude variation $\sigma_A^2$ and (b) fixed $\sigma_A^2$=0 and phase variation $\Delta \phi$.}
\label{fig:beam_squint_variation}
\end{figure}
Fig.~\ref{fig:beam_squint_variation} shows the effect of amplitude and phase errors on the beam pointing accuracy. 
The results show that the squint error remains relatively small compared with the RCS NMSE variation. This is expected since the amplitude and phase errors are modeled as random element-wise perturbations with zero-mean phase error and unit-mean amplitude response. Such random errors mainly reduce coherent gain and distort sidelobes, but they do not introduce a strong systematic phase gradient across the RIS aperture. Therefore, the main effect is RCS degradation rather than a large beam shift. 
\begin{figure}[htbp]
\centering
\includegraphics[scale=0.22]{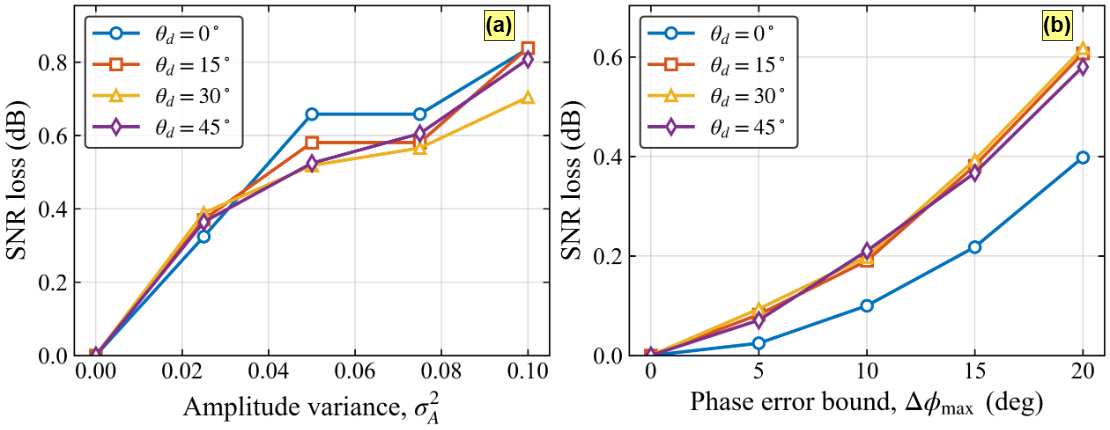}
\caption{SNR loss of the $[16\times10]$ 1-bit RIS for $\theta_i=0^\circ$
and different desired angles $\theta_d$ under (a) amplitude and (b) phase
variations.}
\label{fig:snr_amp_phase_variation}
\end{figure}

Fig.~\ref{fig:snr_amp_phase_variation} shows the SNR loss of the $[16\times10]$ 1-bit RIS for $\theta_i=0^\circ$ under amplitude and phase variations. For each value of $\sigma_A^2$ or $\Delta\phi_{\max}$, the SNR is calculated from the forward and reverse RCS samples and compared with the reference 1-bit RIS case with $A_{mn}=1$ and $\epsilon_{mn}=0$. The SNR loss increases with both amplitude variance $\sigma_A^2$ and phase-error bound $\Delta\phi_{\max}$, showing that amplitude and phase impairments reduce the received signal strength.

\section{Full Wave Simulation}
\label{sec:Full_Wave_Simulation}
\begin{figure}[htbp]
\centering
\includegraphics[width=85mm,height=40mm]{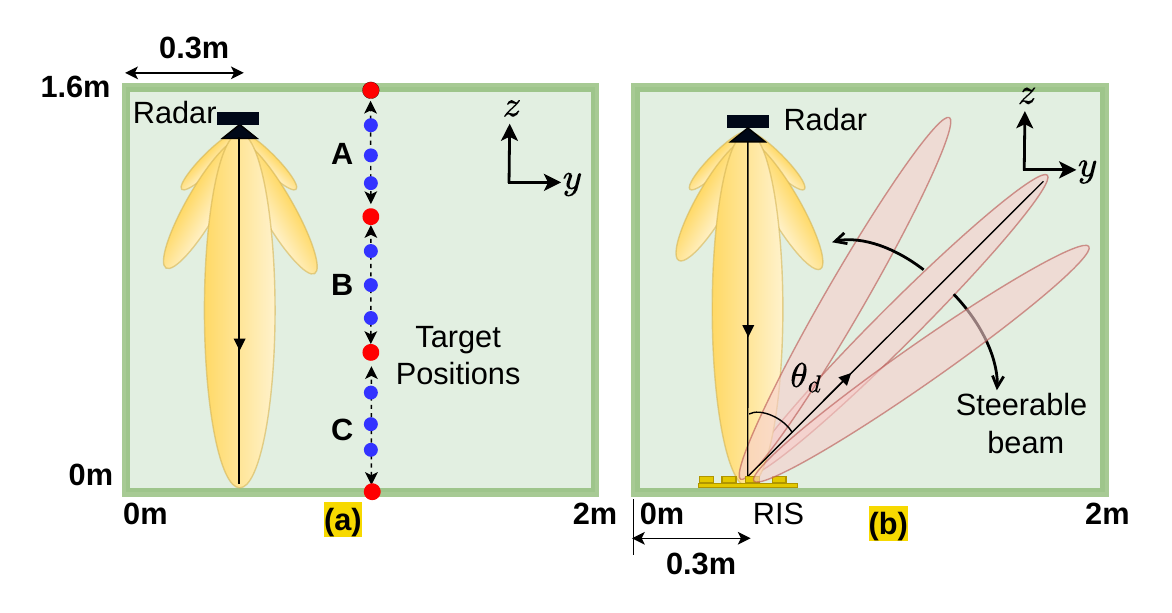}
\caption{Simulation setup for radar system (a) without RIS and (b) with RIS.}
\label{fig:simulation_setup}
\end{figure}

In this section, we consider full-wave electromagnetic simulations of the radar detection performance enhanced by an RIS using CST Microwave Studio's time-domain solver. We consider a 2 m$\times$1.6 m two-dimensional simulation space, along the $yz$ plane as shown in Fig.\ref{fig:simulation_setup}, with a monostatic radar's horn antenna located at $(0.3,1.6)$m with a 3-dB beamwidth of $28.9^\circ$, directed towards the $-z$ axis.
We consider a two-dimensional RIS in the $xy$ plane with identical 16 mm $\times$ 16 mm unit cells tuned for a center frequency of 5.5 GHz. The unit cell design is described in detail in the following section. The RIS is centered at $(0,0.3,0)$  m. The unit cells of the RIS are tuned for specific combinations of incident and desired angles of reflection ($\theta_i;\theta_d$). The full-wave solver models the RIS's realistic performance, accounting for mutual coupling between unit cells, diffraction at the edges, and dielectric and ohmic losses within the unit cells.
The complex vector electric-field distribution is extracted from the software over the beam-steering $yz$ plane ($x$-polarized), and sampled at a 10 mm $\times$ 10 mm spatial grid. To compare the performance enhancement introduced by the RIS, the electric field is simulated in its absence. 

\begin{figure}[htbp]
\centering
\includegraphics[width=85mm,height=50mm]{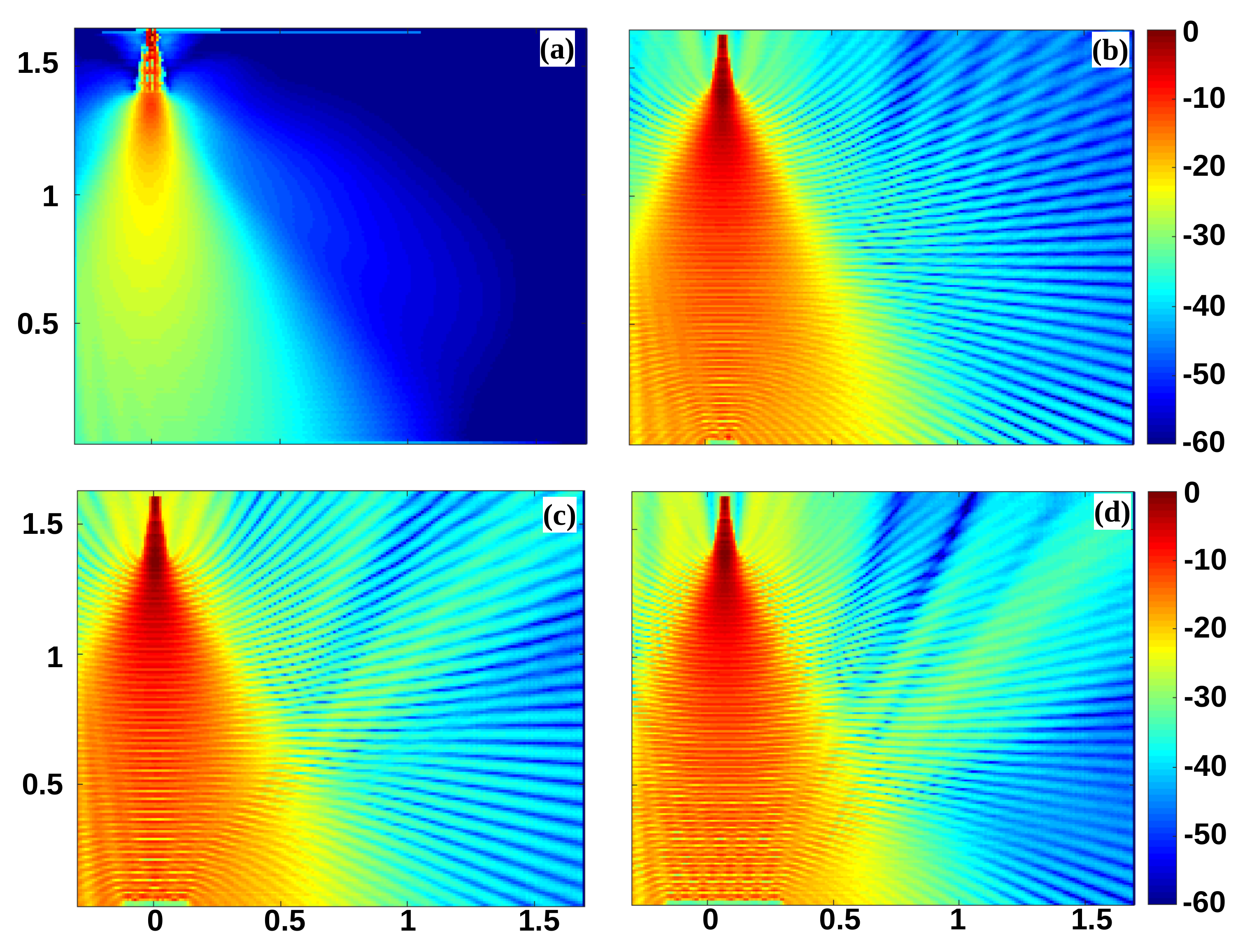}
\caption{Magnitude of electric-field over the $2 \times 1.6~\mathrm{m}^2$ simulation domain for radar system (a) without RIS, and with (b) $[8\times10]$, (c) $[16\times10]$, and (d) $[32\times10]$ unit-cell RIS tuned for ($\theta_i=0^\circ;\theta_d=45^\circ$).}
\label{fig:simulated_efield}
\end{figure}

Figure~\ref{fig:simulated_efield} shows the electric-field magnitude distribution over the simulation domain. In the absence of the RIS, as seen in Fig.~\ref{fig:simulated_efield}(a), the total electric field is primarily dominated by the direct radiation from the horn antenna, resulting in a 
smooth field distribution with spatial focusing based on the beamwidth of the horn antenna and its sidelobe level of -25.6 dB. Next, we consider three independent scenarios with RIS aperture sizes: $[8\times10]$, $[16\times10]$, and $[32\times10]$ along $y$ and $x$ axes, in Figs~\ref{fig:simulated_efield}(b)-(d), respectively. In each case, the RIS is tuned for $(\theta_i = 0^{\circ};\theta_d = 45^{\circ})$. 
The RIS alters the electric field distribution in regions previously outside the radar antenna's field of view, as shown in the figures. 
We observe a strong electric field in regions specifically along the $\theta_d$. The interference between the incident field from the horn and the scattered field from the RIS results in fading patterns in the total field distribution. 
The beamwidth and strength of the scattered signals vary based on RIS aperture size.

\begin{figure}[htbp]
\centering
 \includegraphics[scale=0.25]{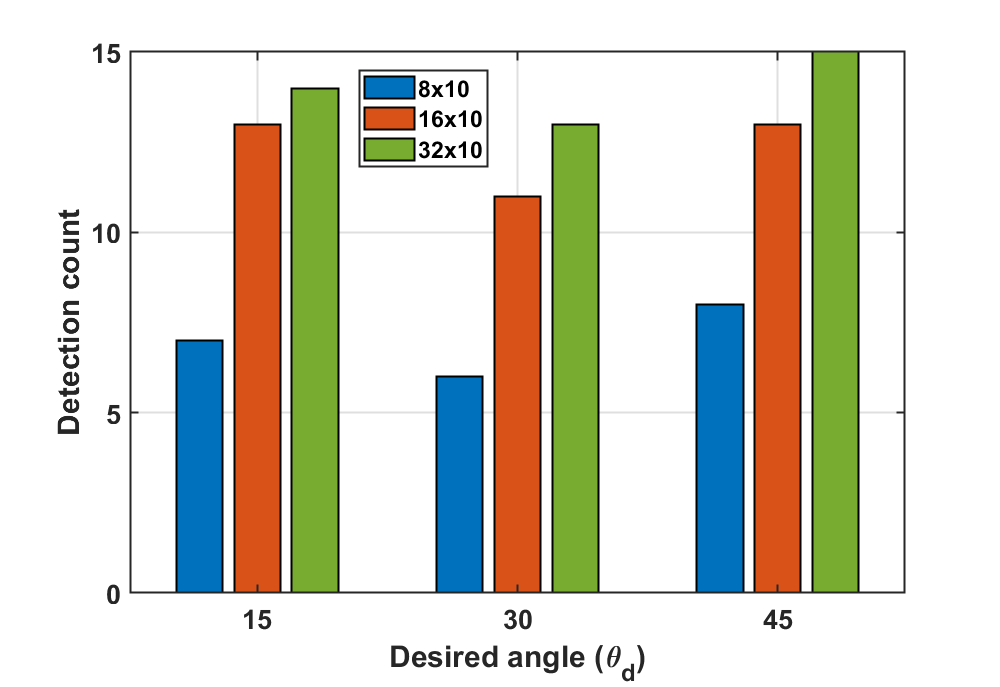}
\caption{Detection performance for different RIS sizes ($[8\times10]$, $[16\times10]$, and $[32\times10]$) under varying $\theta_d$.}
\label{fig:detection_comparison}
\end{figure}

We consider a radar target modeled as a single isotropic point scatterer located at $y = 1$ m, and along 50 different positions spanning sets: \{A\}, \{B\}, and \{C\}, spaced 30 mm apart along the $z$ axis, as shown in Fig.~\ref{fig:simulation_setup}. All the target points between $z=1.0$ m and $z=1.6$ m belong to set \{A\} (15 positions), all the points between $z=0.6$ m and $z=1$ m belong to set \{B\} (17 positions), and all the remaining target positions belong to set \{C\} (18 positions). 
For each target position belonging to sets \{A\}, \{B\}, or \{C\}, the electric field distribution is simulated for RIS configured for $\theta_d=45^{\circ},30^{\circ},15^{\circ}$ respectively, while $\theta_i=0^{\circ}$ for all three cases. In other words, we assume that the RIS is tuned to direct the mainlobe along one of the three $\theta_d$ values nearest the target point.
At each position, the electric-field magnitude at the nearest grid coordinate is sampled directly from the full-wave simulated field. 
The electric field values are subsequently used to compute the SNR at the radar as per \eqref{eq:snr}, assuming the target RCS is $ 1 m^2$, and the noise is fixed at $0$ dBm. The target is assumed to be detected when the SNR is above 1 dB. \\
\indent In the absence of an RIS, the system detects the target at only \emph{one of 50} positions, indicating very limited sensing capability of the target outside the mainlobe of the radar antenna (assuming the radar's antenna is not capable of scanning). The detection performances of the different RIS aperture sizes: $[8\times10]$ (blue bar graph), $[16\times10]$ (brown bar graph), and $[32\times10]$ (green bar graph), are compared in Fig.~\ref{fig:detection_comparison}. 
The figure shows the number of successful detections per set for each $\theta_d$ corresponding to $\{15^{\circ},30^{\circ},45^{\circ}\}$. We observe that as the RIS size increases, the number of detections per desired angle also increases. With a $[32\times10]$ RIS aperture size, the target is detected in a total of 41 of the 50 total positions by using programmable beam steering of the RIS across the three angles of $\theta_d$.
These results highlight that despite the limitations of quantization, mutual coupling, diffraction, and bi-directional path loss metrics introduced by a finite-sized realistic lossy RIS, there is a substantial increase in the overall detection metrics of a target using RIS-enhanced radar. 
\section{Measurements}
\label{sec:measurement}
This section presents experiments conducted with a hardware prototype of a $[16\times10]$ 1-bit-quantized RIS to assess its performance in enhancing radar detection. All experiments are carried out in a 4 m $\times$ 4 m anechoic chamber, which provides a reflection-free, electromagnetically isolated environment to ensure accurate and reproducible measurements. We measure both $\sigma_f$ and $\sigma_b$  to quantify their reflective behavior as a function of ($\theta_i;\theta_d$). These measurements also provide the beam-squint error and PSLR. Finally, we electronically adjust the RIS phase configuration to direct radar signals toward targets outside the radar's mainlobe and capture their reflected signals to quantify detection performance. 
\subsection{Hardware Prototype of 1-bit RIS}
\begin{figure}[htbp]
\centering
\includegraphics[scale =0.35]{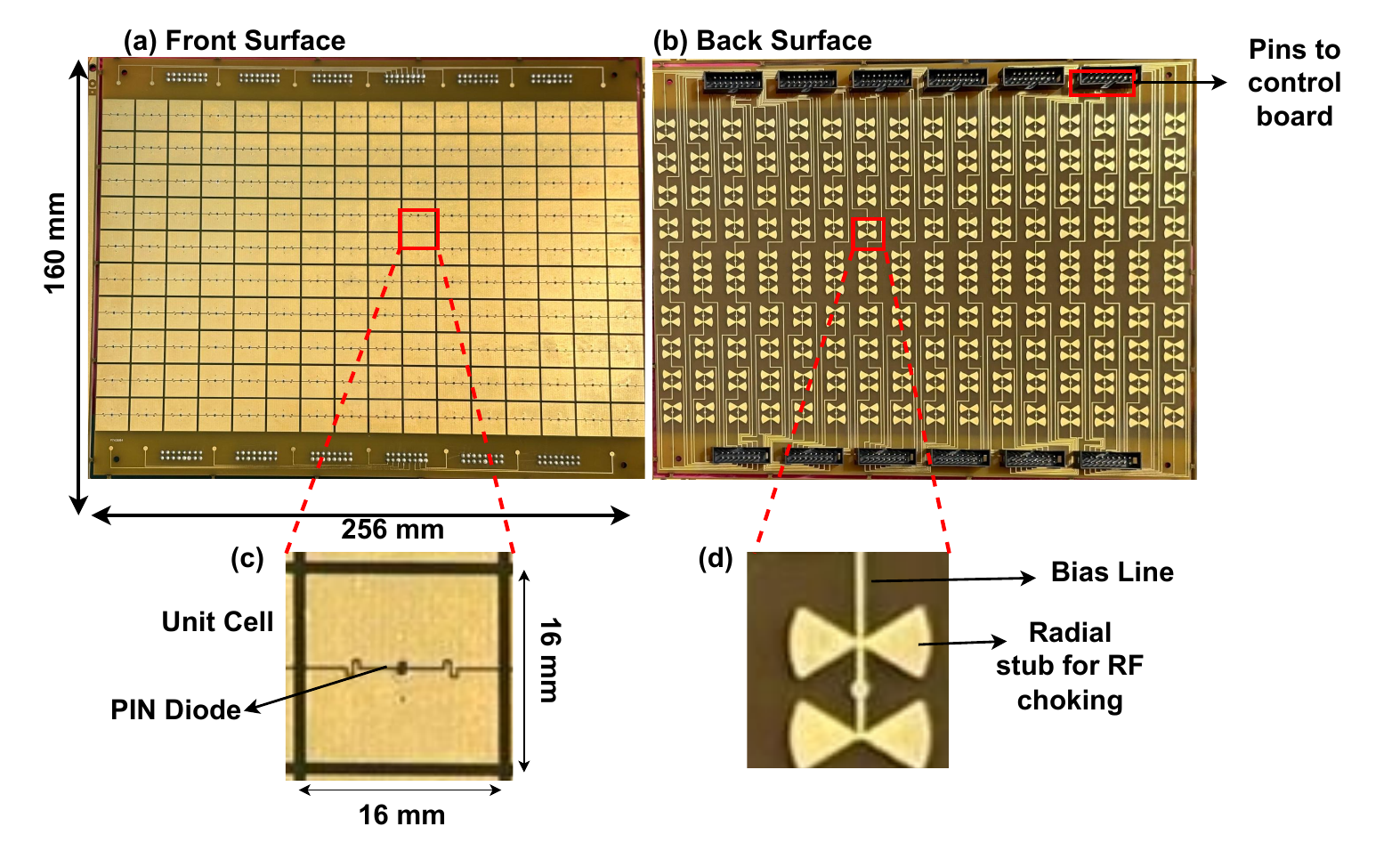}
\caption{Fabricated 1-bit $[16\times10]$ RIS operating at 5.5 GHz: (a) Front view showing periodic array of unit cells, (b) Back view illustrating biasing and control lines. Zoomed view of the unit cell (c) Front view and (d) Back view.}
\label{fig:fabricated_RIS}
\end{figure}

The fabricated RIS panel is shown in Fig.~\ref{fig:fabricated_RIS}, with front and back views depicted in Figs.~\ref{fig:fabricated_RIS}a and \ref{fig:fabricated_RIS}b, respectively. Each unit cell of the RIS is designed as a single metasurface element capable of realizing both the \emph{0} and \emph{1} states of a 1-bit coding metasurface. Digital switching between these two states is achieved through the integration of a PIN diode within each unit cell, as illustrated in Fig.~\ref{fig:fabricated_RIS}.
Each unit cell comprises three metal layers that sandwich two dielectric layers. The top metal layer is a wide dipole loaded with a PIN diode. The middle metal layer serves as the ground, and the bottom metal layer is used for bias lines. The two dielectric layers have thicknesses of 1.6 mm and 0.8 mm, respectively, and are fabricated from FR4 ($\epsilon_r$ =\ 4.3 and a loss tangent of 0.02). The overall dimensions of each unit cell are 16 mm $\times$ 16 mm.

\begin{figure}[htbp]
\centering
\includegraphics[width=95mm,height=42mm]{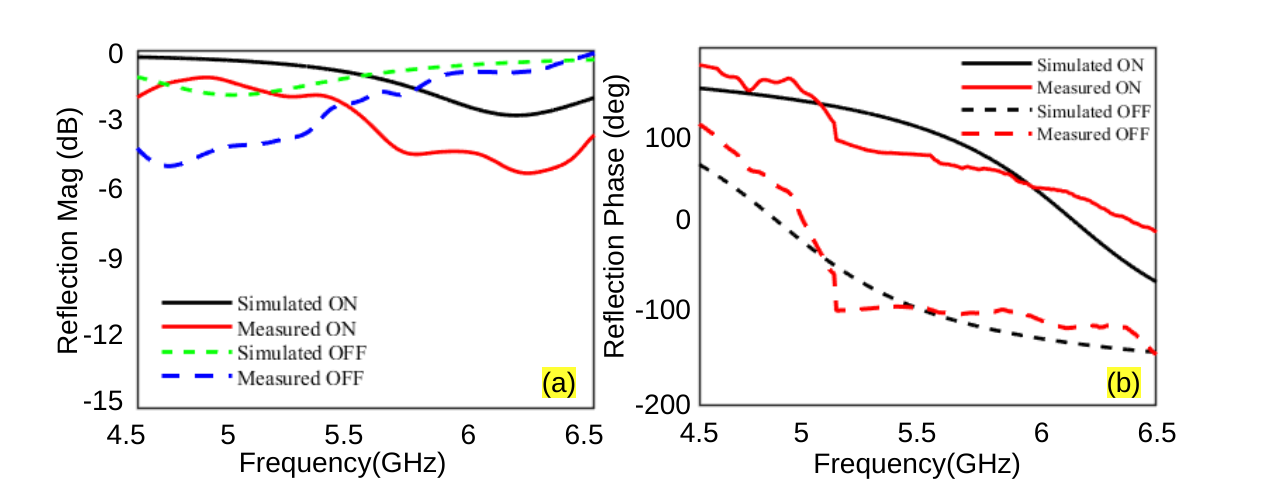}
\caption{Reflection coefficient of unit cell: (a) Magnitude response and (b) Phase response.}
\label{fig:unit_cell_reflection}
\end{figure}

When a forward-bias voltage of 0.9 V is applied, the diode conducts and forms a low-impedance path (\emph{ON} state), thereby modifying the surface current distribution and resulting in a reflection phase.  When no bias (i.e., 0 V) is applied (\emph{OFF} state), the diode exhibits a high impedance and behaves as a small capacitance, yielding a different reflection phase. A zoomed-in view in the inset of Fig.~\ref{fig:fabricated_RIS} illustrates the PIN diode connection, biasing lines, and radial stubs used for radio frequency and direct current (DC) isolation. The DC bias lines are connected to the PIN diodes through vias, allowing a control unit to selectively switch some cells ON and others OFF, thereby dynamically reconfiguring the metasurface's reflection pattern. The full-wave simulated and measured magnitude and phase of the reflection coefficient of the unit cell are presented in Fig.\ref{fig:unit_cell_reflection}. We observe that the fabricated metasurface exhibits a wide fractional bandwidth of 18.51\% centered at 5.5 GHz, where the phase difference remains within $180^\circ \pm 20^\circ$. The simulated and measured reflection loss at 5.5 GHz are below 1.0 dB and 2.3 dB, respectively.
\subsection{RCS Pattern Measurement} 
\subsubsection{Experimental Setup}
\begin{figure}[htbp]
\centering
\includegraphics[width=90mm,height=72mm]{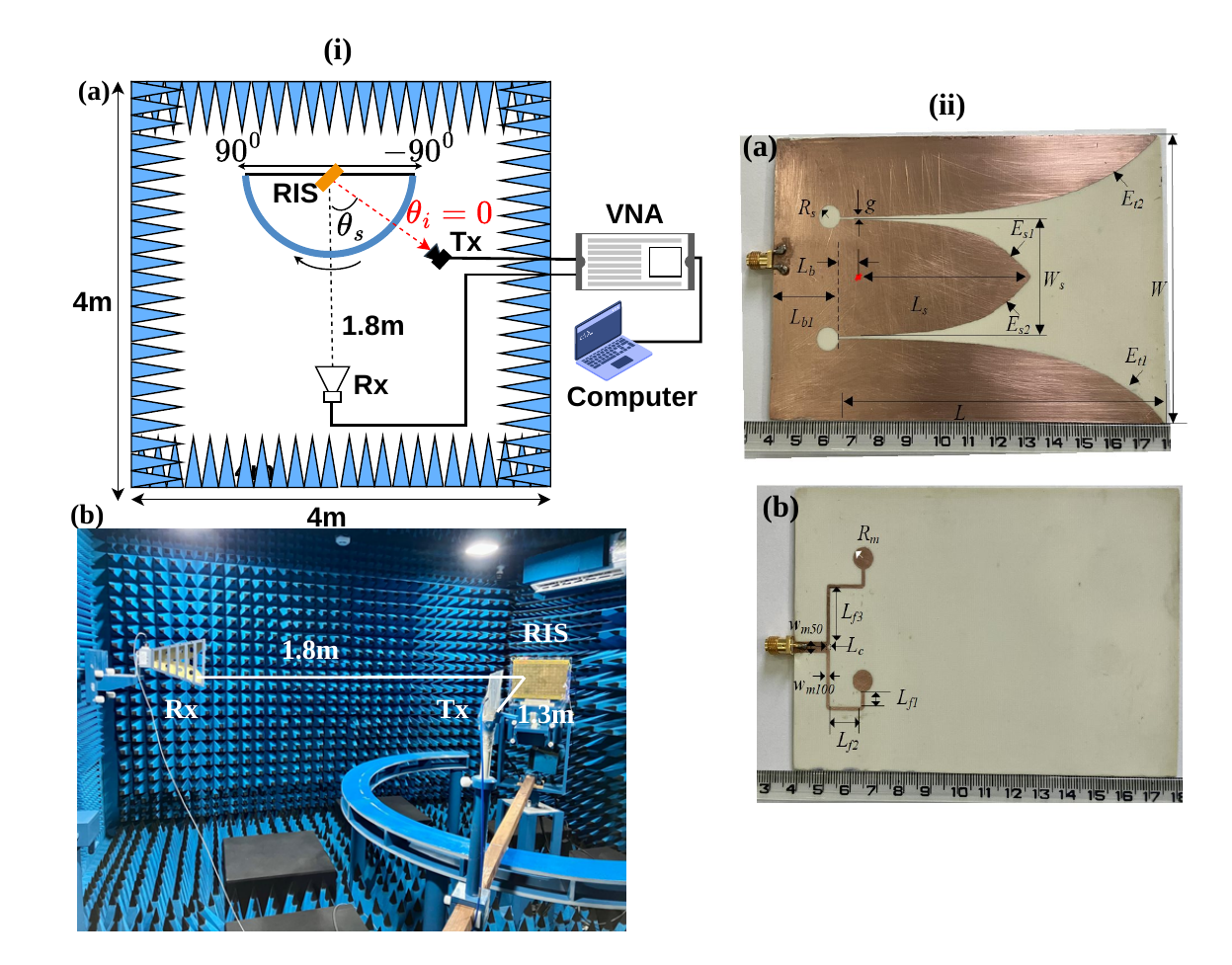}
\caption{(i) (a) Top view of the scattering pattern measurement setup inside an anechoic chamber for the 1-bit $[16\times10]$ RIS for ($\theta_i=0^{\circ};\theta_d$). (b) Photograph of the experimental setup. (ii) Photograph of the custom-made Vivaldi antenna used as a transmitter for the measurement of the scattering field pattern of the RIS inside the anechoic chamber. (a) top view, (b) backside view.}
\label{fig:exp1}
\end{figure}

Our first set of experiments measures the scattering pattern of RIS using the setup shown in Fig.~\ref{fig:exp1}. Here, a custom-made Vivaldi antenna is used as a transmitting (Tx) antenna and connected to the first port of a Keysight E5063A vector network analyzer (VNA). A receiving horn antenna (Rx), LB-8180-SF, operates at a frequency of 5.5 GHz with a gain of 10.7 dB and $45.9^{\circ}$ beamwidth, and is connected to the VNA's second port. Each antenna is positioned so that it lies entirely outside the main lobe of the other antenna. The fabricated RIS, described previously, is placed before both antennas, as shown in the figure, at distances of 1.3 m and 1.8 m from the Vivaldi and horn antennas, respectively. Transmission parameters, $S_{21}$, are measured using the VNA. Since the anechoic chamber absorbs all multipath from the walls and floors, the primary signal at the receiver antenna is the direct signal reflected by the RIS. 
The VNA is configured to operate at a center frequency of 5.5 GHz, intermediate-frequency bandwidth of 1 kHz, and a frequency sweep of 1601 points from 5495 MHz to 5505 MHz, i.e., a 10 MHz bandwidth. The transmitted power level is set to the maximum available output of 0 dBm. The experiments are performed with the transmitting antenna directed perpendicular to the RIS surface ($\theta_i=0^{\circ}$), where the gain of the transmitting antenna is maximum. Typically, the main lobe of the scattered signal from a regular metal plane will be along the specular angle with respect to the incident wave. However, an RIS can be configured to direct the mainlobe at a non-specular, desired angle of reflection $\theta_d$. To study the scattering pattern from the RIS, we fixed the RIS with the transmitting antenna on a motorized turntable, remotely controlled via a computer in a dedicated control room adjacent to the chamber, as shown in \ref{fig:exp1}(i). The measurement ensures that the incident angle, $\theta_i=0^{\circ}$, remains fixed as the turntable rotates, but the RCS pattern is captured for varying scattered angle, $\theta_s$. The turntable rotates with a step size of $1.8^\circ$ from $-90^\circ$ to $90^\circ$ as shown in Fig.~\ref{fig:exp1}a. We conduct experiments for three RIS configurations: $\theta_d= \{15^{\circ}, 30^{\circ}, 45^{\circ}\}$.

We designed and fabricated a custom Vivaldi antenna for use as a transmitter in these experiments for the following reasons. First, the arm attached to the turntable for mounting the transmitting antenna cannot withstand the load of the heavy broadband horn antenna (used in other experiments) during multiple rotations. Instead, it was safer and easier to use the planar and lightweight Vivaldi antenna. Second, as the transmitter rotates in front of the receiving antenna, the Vivaldi antenna provides less feed blockage. The transmitter is placed 1.3 m away from the RIS, to ensure far-field incidence conditions.
The top and bottom views of the Vivaldi antenna are shown in Fig.~\ref{fig:exp1}(ii). The figure illustrates the geometry of the Vivaldi antenna, including its tapered slot structure and feeding network. The slot profile is defined using exponential tapering to ensure a smooth transition from the feed gap $g$ to the slot width $W_s$. The inner edges $(E_{s1}, E_{s2})$ and outer edges $(E_{t1}, E_{t2})$ are symmetrically described by exponential functions using equations \ref{eq:vivaldi_eq1} and \ref{eq:vivaldi_eq2}, providing a gradual impedance transformation along the length of the structure.
\begin{equation}
\small
\begin{aligned}
E_{s1} / E_{s2} = \pm \frac{1}{2} \left[
W_s - g\exp \left( \ln \left( \frac{W_s}{g} \right)\frac{x}{L_s} \right)\right],\\
\quad \text{for } 0 \le x \le L_s
\end{aligned}
\label{eq:vivaldi_eq1}
\end{equation}
\begin{equation}
\small
\begin{aligned}
E_{t1} / E_{t2} = \pm \frac{1}{2} \left[
W_s + g\exp \left( \ln \left( \frac{W - W_s}{g} \right)\frac{x}{L_s} \right)
\right], \\
\quad \text{for } 0 \le x \le L  
\end{aligned}
\label{eq:vivaldi_eq2}
\end{equation}
This tapering minimizes reflections and supports broadband operation. The indicated dimensions represent the key design parameters governing the slot expansion and feeding mechanism. The antenna is designed on a Rogers RO4350B substrate with $\varepsilon_r = 3.66$, $\tan \delta = 0.0037$, and thickness $h$ = 1.524 mm. The design parameters, indicated in the figure, in millimeters are $W = 95$, $L = 110$, $W_s = 40$, $L_s = 60$, $g = 0.9$, $R_m = 3.5$, $R_s = 4$, $L_b = 5.8$, $L_{b1} = 25$, $L_{f1} = 5.5$, $L_{f2} = 12$, $L_{f3} = 19.2$, $W_{m50} = 3.33$, $W_{m100} = 0.83$, and $L_c = 2$. 
The antenna's operating bandwidth is $4$--$8.5$~GHz.

\begin{figure}[!t]
    \centering
    \includegraphics[scale=0.2]{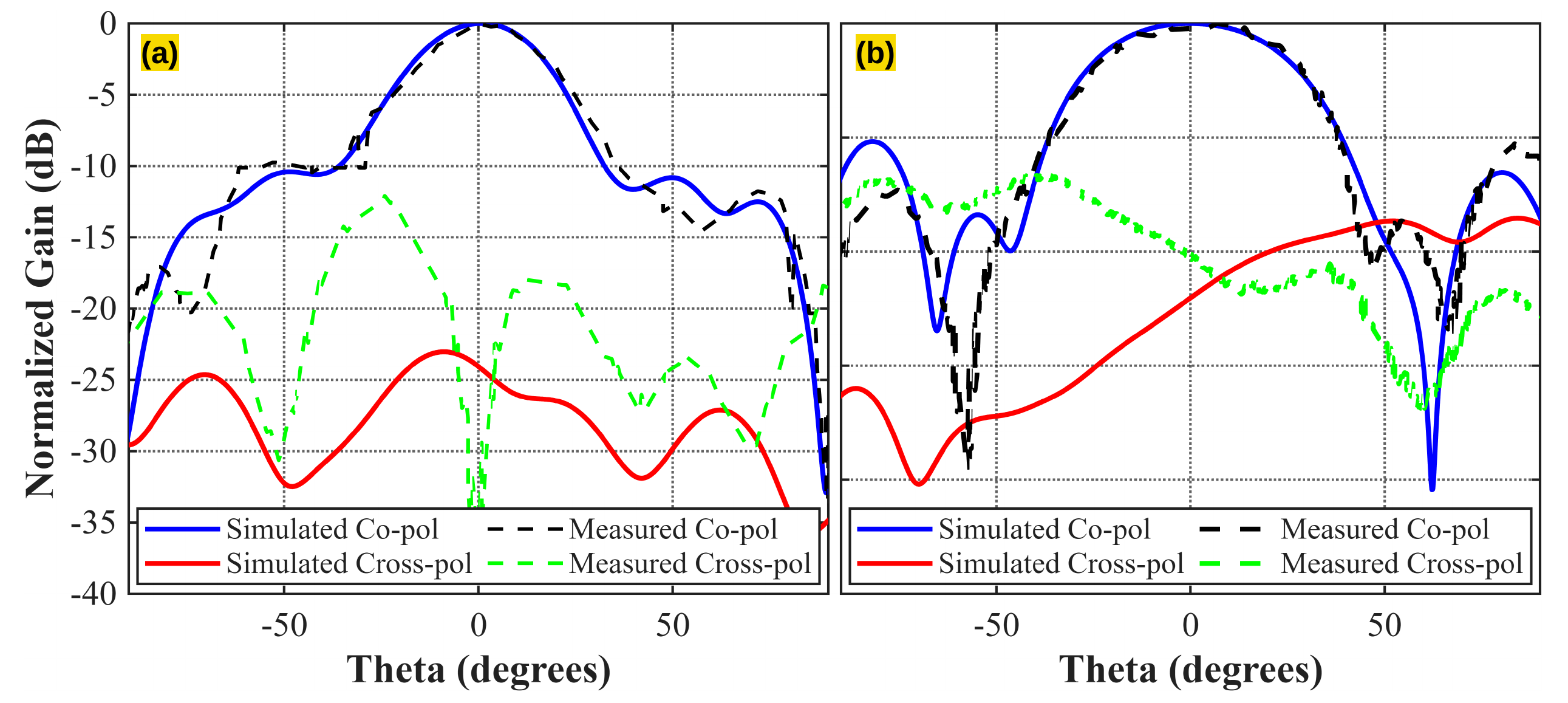}   
    \caption{Simulated and measured normalized radiation patterns of the Vivaldi antenna at 5.5 GHz in (a) $xz$-plane and (b) $yz$ plane.}
    \label{fig:vivaldi_radiation_pattern}
\end{figure}

The simulated and measured radiation patterns of the Vivaldi antenna at 5.5~GHz for both the $xz$- and $yz$-planes are shown in Fig.~\ref{fig:vivaldi_radiation_pattern}. It can be observed that the radiation patterns in both the $xz$- and $yz$-planes are quite symmetric. Moreover, in the $xz$-plane, the cross-pol level and side lobe level (SLL) are $17.08$~dB and $10.35$~dB below the main lobe maximum level, respectively. In the $yz$-plane, the corresponding cross-pol level and SLL are $23.03$~dB and $10.41$~dB below the main lobe maxima, respectively.
\subsubsection{Results}
\begin{figure*}[htbp]
\centering
\includegraphics[scale=0.3]{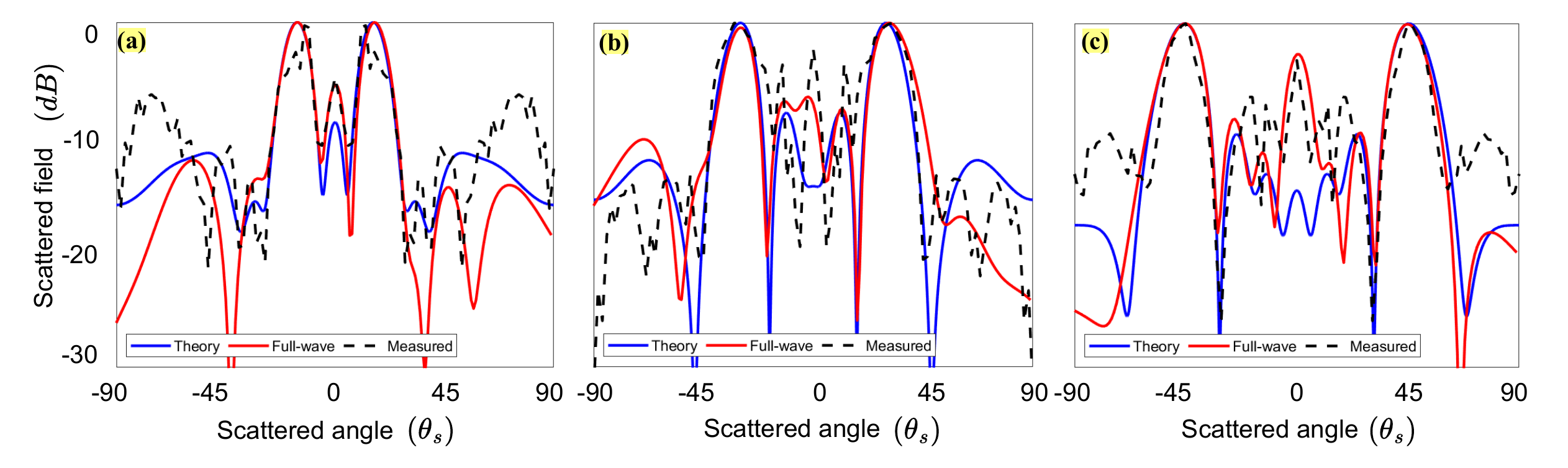}
\caption{Normalized radiation pattern obtained from theory, full-wave simulation, and measurement for $\theta_d$ at (a) $15^\circ$, (b) $30^\circ$, and (c) $45^\circ$.}
\label{fig:norm_Es_sim_exp}
\end{figure*}

The measured scattering patterns for $\theta_i=0^{\circ}$ are presented in Fig.~\ref{fig:norm_Es_sim_exp}. 
The normalized measurement results (dotted black lines) are compared with the scattering patterns obtained using the theory presented in the section.\ref{sec:theory} (solid blue lines) and from simulations using the full-wave solver in Section.\ref{sec:Full_Wave_Simulation} (solid red lines). In Fig.\ref{fig:norm_Es_sim_exp}a, the RIS is configured for $\theta_d = 15^{\circ}$. The theoretical, simulation, and measured results show the main lobe at 15$^{\circ}$. Due to the one-bit quantization, we observe an additional grating lobe at -15$^{\circ}$. We observe similar results when the RIS is tuned for $\theta_d = 30^{\circ}$ and $\theta_d = 45^{\circ}$ for the same incident angle in Fig.\ref{fig:norm_Es_sim_exp}b and c. The theoretical results accurately reflect the RIS's aperture response. 
To quantify the performance, we present the beam squint error, $\left| \theta_d' - \theta_d \right|$, between the desired beam direction ($\theta_d$) and the actual main lobe beam position estimated from the theoretical, simulated, and measured beam patterns in the figure ($\theta_d'$) in Table.\ref{table:beam_squint_errors}.
\begin{table}[h!]
\centering
\caption{Beam squint errors for a 1-bit $[16\times10]$ RIS at $(\theta_i=0^{\circ};\theta_d)$, based on theory, simulations, and measurements.}
\label{table:beam_squint_errors}
\begin{tabular}{c|cc|cc|cc}
\hline\hline
\multirow{2}{*}{$\theta_d (^{\circ})$} & \multicolumn{2}{c|}{Theory} & \multicolumn{2}{c|}{Simulation} & \multicolumn{2}{c}{Measurement} \\
 & $\theta_d' (^{\circ})$ & Error & $\theta_d' (^{\circ})$  & Error  & $\theta_d' (^{\circ})$ & Error \\
\hline 
$ 15$ &  15.3&  0.3&  15.1&  0.1& 17.5 & 2.5  \\
$ 30$ &  29.6&  0.4&  30.1&0.1  & 31 & 1 \\
$ 45$ &  44.7&  0.3&  44.8& 0.2 & 46 & 1 \\
\hline\hline
\end{tabular}
\end{table}
The theoretical and simulated results agree well, with angular discrepancies of no more than $0.4^\circ$, indicating that the analytical model accurately predicts the RIS response under ideal conditions. In contrast, the measured results show larger deviations, particularly at smaller steering angles. For instance, at $\theta_d = 15^\circ$, the measured beam is observed at $17.5^\circ$, resulting in a $2.5^\circ$ offset, whereas the theoretical and simulated errors are only $0.3^\circ$ and $0.1^\circ$, respectively. The deviation decreases at larger steering angles, with measured errors of approximately $1^\circ$ for $30^\circ$ and $45^\circ$ targets.
The observed errors in the measured angles indicate that the practical RIS introduces additional errors not captured by the theoretical or simulated models. These differences are mainly due to the calibration errors in the measurement setup. Nevertheless, the overall beam-steering behavior remains consistent among theory, simulation, and experiment, confirming the proper operation of the 1-bit RIS.

The second metric is PSLR, which, in this case, is measured for specular reflection angle at 0$^{\circ}$. The PSLR for all four cases is summarized in Table~\ref {table:pslr}. The results indicate that the RIS can successfully direct the strongest reflection beam to the desired angle rather than the specular reflection angle.
\begin{table}[h!]
\centering
\caption{PSLR (dB) for a 1-bit $[16\times10]$ RIS at ($\theta_i=0^{\circ},\theta_d$), based on theory(ideal), theory(actual), simulations, and measurements.}
\label{table:pslr}

\begin{tabular}{p{0.8cm}|p{1.2cm}|p{1.4cm}|p{1.2cm}|p{1.5cm}}
\hline
$\theta_d(^\circ)$ & Theory ideal (dB) & Theory actual (dB) & Simulations (dB) & Measurements (dB) \\
\hline
$ 15$ &8.6  &4.8 &3.4  &4.8 \\
$ 30$ &14.2 &9.7 &3.0 &2.3  \\
$ 45$ &14.5 &0.2 &1.1  &2.3 \\
\hline
\end{tabular}
\end{table}
The theoretical PSLR is above the simulated and measured values. The relatively small degradation in the measured PSLR demonstrates good agreement with predictions and effective suppression of the specular reflection at small steering angles. However, actual 1-bit PSLR shows good agreement with measurement. Larger discrepancies are observed at higher steering angles arising from phase-quantization errors inherent to the 1-bit RIS, as well as increased sensitivity of the ON and OFF state reflection magnitude and phase values to fabrication tolerances and finite-aperture effects.

\subsection{Radar Cross-Section of RIS}
\subsubsection{Experimental setup}
\begin{figure}[htbp]
\centering
\includegraphics[width=90mm,height=45mm,keepaspectratio]{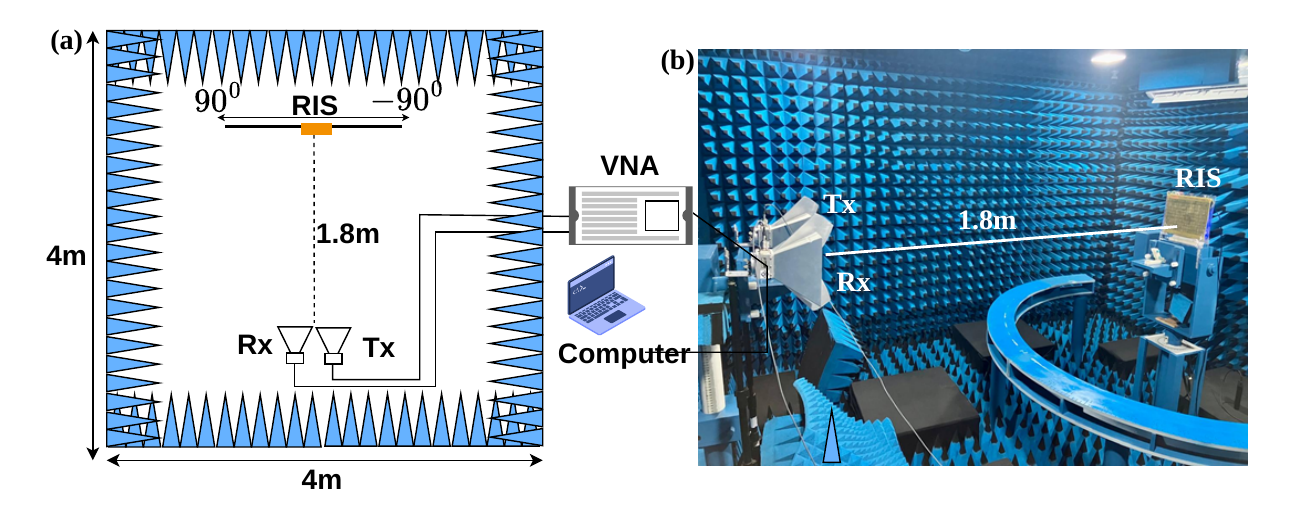}
\caption{(a) Top view of the RCS measurement setup inside an anechoic chamber for the 1-bit $[16\times10]$ RIS for ($\theta_i=0^{\circ};\theta_d$); (b) Photograph of the experimental setup.}
\label{fig:exp2}
\end{figure}

In our second experiment, we configure the two antennas and the VNA as a monostatic narrowband radar as shown in Fig.~\ref{fig:exp2}a. Here, the transmitting and receiving antennas are in proximity while the RIS is at a distance of 1.8 m before them in the far-field, as shown in Fig.~\ref{fig:exp2}b. The VNA is operated at 5.5 GHz with a 1 kHz bandwidth and is set to transmit at 0 dBm. We used HF907 horn antennas, operated at 5.5 GHz, with a gain of 10 dBi. Based on the measured $S_{21}$, the RCS of RIS is calculated using Friis's radar equation \cite{balanis2024balanis}.
\subsubsection{Results}
Table \ref{table:RCS_of_RIS_monostatic} presents a quantitative comparison of RCS values in dBsm obtained from theory, full-wave simulations, and measurements. For all cases, the incident angle is fixed at $\theta_i = 0^\circ$, while $\theta_d$ is varied. When $\theta_d = 0^\circ$, the theory predicts an RCS of 8.5 dBsm, whereas the full-wave simulation yields a lower value of 4.8 dBsm, resulting in a difference of 3.7 dB. The measured RCS is significantly lower at –2.5 dBsm, corresponding to deviations of 11.0 dB and 7.3 dB from the theoretical and simulated results, respectively. This large discrepancy can be attributed to idealized assumptions in the theoretical model (e.g., no edge effects, lossless dielectrics and conductors, and perfect phase control). Losses, edge effect, and mutual coupling are captured in simulation but not in theory, while additional factors, such as fabrication tolerances, surface roughness, and calibration errors, affect the measurement results.

For higher angles of $\theta_d$, the theoretical RCS is closer to the simulated and measurement values. These discrepancies indicate that the theoretical model tends to overestimate the beam-steering degradation at larger reflection angles, whereas simulations and measurements capture additional scattering contributions and edge diffraction effects that partially enhance the RCS.
\begin{table}[htbp]
\centering
\caption{Monostatic RCS of 1-bit $[16\times10]$ RIS from theory, simulation, and measurement.}
\begin{tabular}{c|c|c|c|c}
\hline
$\theta_i(^{\circ})$ & $\theta_d(^{\circ})$ & Theory (dB) & simulation (dB)& Measured (dB) \\
\hline\hline
0 &  0 &  8.5 & 4.8  & -2.5\\
0 & 15 & -3.5 & -4.0 & -6.6\\
0 & 30 & -9.5 & -5.7 & -6.6\\
0 & 45 & -9.6 & -5.7 & -7.1\\
\hline \hline
\end{tabular}
\label{table:RCS_of_RIS_monostatic}
\end{table}

For the largest reflection angle, $\theta_d = 45^\circ$, the theoretical RCS is –9.6 dBsm and the simulated value remains at –5.7 dBsm, yielding a difference of 3.9 dB. The measured RCS further decreases to –7.1 dBsm, which is 2.5 dB lower than the simulated result and 2.5 dB higher than the theoretical prediction. This is due to increased phase errors, stronger edge diffraction, and reduced effective aperture at large steering angles, which are not fully captured by the simplified theoretical model but are accounted for in full-wave simulations and experimental observations.
Overall, the comparison demonstrates consistent trends across theory, simulation, and measurement: increasing $\theta_d$ reduces RCS. The remaining quantitative differences are primarily due to idealized theoretical assumptions, numerical approximations in simulations, and unavoidable nonidealities in fabrication and measurement setups.
\subsection{Micro-Dopplers of Humans in Non-specular Regions}
\subsubsection{Experimental Setup}
\begin{figure}[htbp]
\centering
\includegraphics[width=90mm,height=40mm, keepaspectratio]{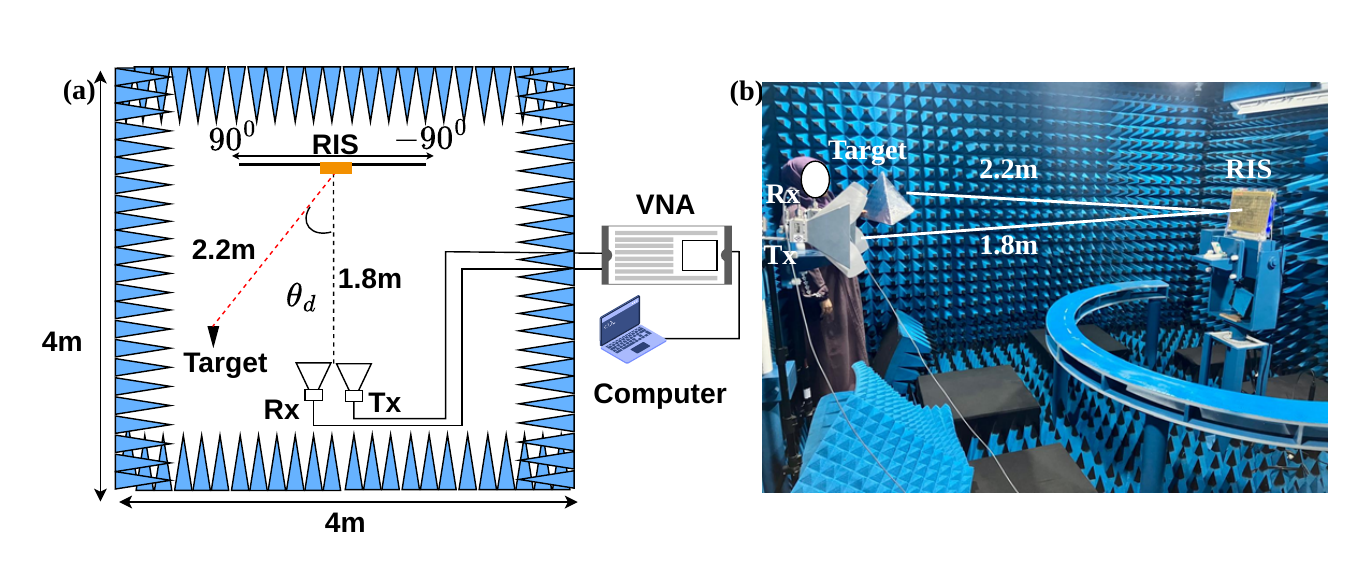}
\caption{(a) Top-view of the RIS-enhanced radar system using a corner reflector inside an anechoic chamber for ($\theta_i = 0^{\circ},\theta_d$). (b) Photograph of the experimental setup.}
\label{fig:exp3}
\end{figure}

In the third experiment, the setup comprises a monostatic radar system, an RIS, and a human subject (with informed consent) holding a 36 cm $\times$ 30 cm (base $\times$ height) trihedral corner reflector serving as the radar target. The monostatic radar is configured with the two horn antennas and VNA as described in the previous experiment, and illustrated in Fig.~\ref{fig:exp3}a. The human is located 2.2 m from the RIS and positioned outside of the main lobe of the radar antennas as shown in Fig.~\ref{fig:exp3}b. Further, the human swings the corner reflector with one hand, generating micro-Doppler returns. The radar and the RIS are separated by 1.8 m. The objective of the experiment is to assess the effectiveness of the RIS in enhancing the radar micro-Doppler signatures of human movements.

The experiments are performed for two scenarios. In the first scenario, the experiments are conducted with the radar and the target in the absence of the RIS. The experiments are then repeated in the presence of the RIS. Note that all other parameters, such as the radar parameters, position of the target with respect to the radar, and orientation of the radar antennas, are not changed across both experiments. The received radar signal is further processed to generate a joint time-frequency spectrogram by applying the short-time Fourier transform (STFT) to the received signal $s(t)$ as shown in
\par\noindent\small
\begin{align}
  \textbf{STFT}(t,f) = \int_{} s(\tau)h(t-\tau)e^{-j2\pi f\tau}d\tau,
\end{align}
\normalsize
where $h(t)$ is the window function with a duration of 0.1 seconds.

\subsubsection{Results}
\begin{figure}[t]
\centering
\includegraphics[width=100mm,height=60mm]{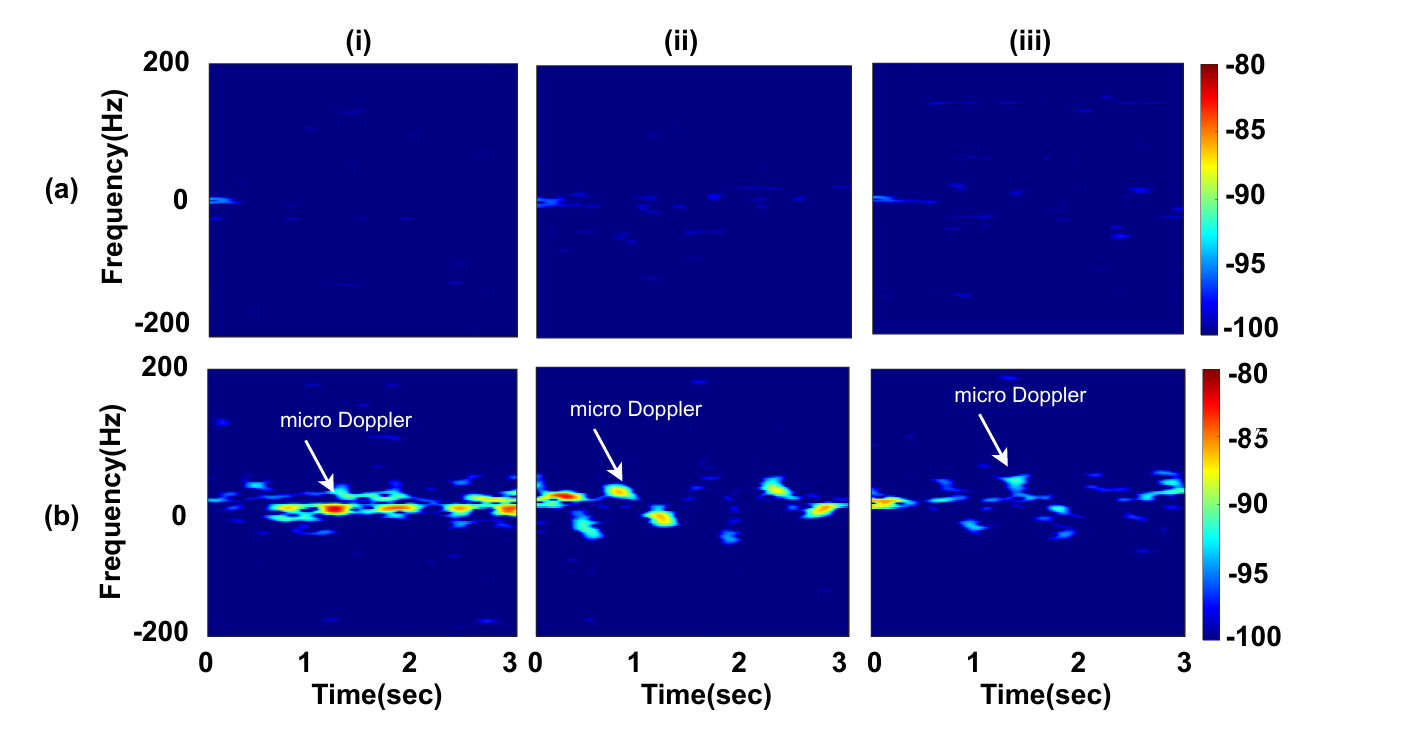}
\caption{Doppler spectrograms generated from measurement data at 5.5 GHz for $\theta_i=0^\circ$.
(a) The top row presents results without RIS, while (b) the bottom row presents results with RIS for $\theta_d$ equal to (i)~$15^\circ$, (ii)~$30^\circ$, and (iii)~$45^\circ$.}
\label{fig:Spectrogram}
\end{figure}

Figure \ref{fig:Spectrogram} illustrates the spectrograms of the received radar signal under different measurement scenarios. The first row corresponds to the configuration without RIS, while the second row shows the RIS-assisted measurements. Each column, labeled (i)--(iii), represents a different configuration of RIS  corresponding to $\theta_d = 15^\circ, 30^\circ$, and $45^\circ$, respectively, with the  $\theta_i = 0^\circ$.
In the top row, without RIS, all spectrograms show no distinct micro-Doppler components. The absence of micro-Doppler is because the human is entirely outside the mainlobe of the radar antenna and, for practical purposes, invisible to the radar due to the low SNR. In contrast, in the second row, micro-Doppler signatures are obtained in all cases and highlighted in the spectrograms. Here, we can clearly discern the Doppler returns from the motion of the trihedral reflector. The Doppler returns are positive when the hand moves towards the radar and negative when it moves away from the radar. Note that we do not observe the Doppler returns from the second hand, which does not hold the trihedral reflector, due to weak returns. Also, the returns from torso, head, and legs fall at zero Doppler and are removed through DC filtering. The Doppler values of the human motion vary in value for different $\theta_d$ due to the changes in bistatic radial velocity. Overall, the results demonstrate that the RIS effectively manipulates electromagnetic-wave propagation, enabling the detection of human motion using hand micro-Doppler signals.

\section{RIS-Assisted Sensing in Real Environment}
\label{sec:real_env}
This section presents the experimental validation of the proposed RIS-assisted sensing approach in a real indoor environment. First, the target detection performance is evaluated using measured data under both RIS-assisted and non-RIS conditions. Next, a measurement repeatability and uncertainty analysis is conducted to assess the reliability and consistency of the experimental results. Finally, region of convergence (ROC)-based detection performance is analyzed using the measured environment data to quantify the detection capability of the proposed system.

\subsection{Target Detection using RIS in NLOS Scenario}
\begin{figure}[htbp]
\centering
\includegraphics[width=67mm,height=60mm]{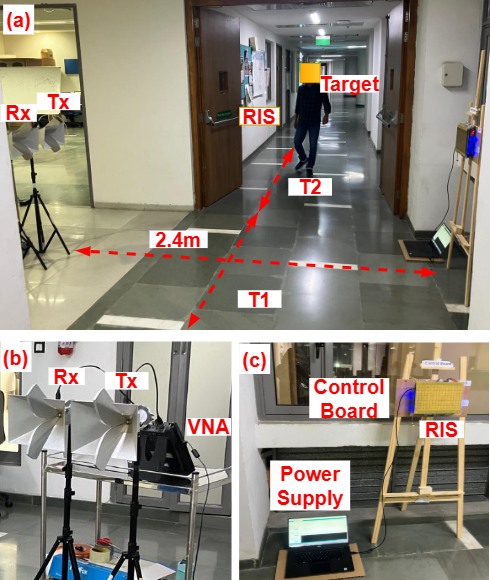}
\caption{(a) Experimental radar setup for target detection in a real environment using RIS (b) close view of radar setup and (c) close-up view of RIS Setup.}
\label{fig:real_exp_setup}
\end{figure}

\begin{figure}[htbp]
\centering
\includegraphics[scale=0.5]{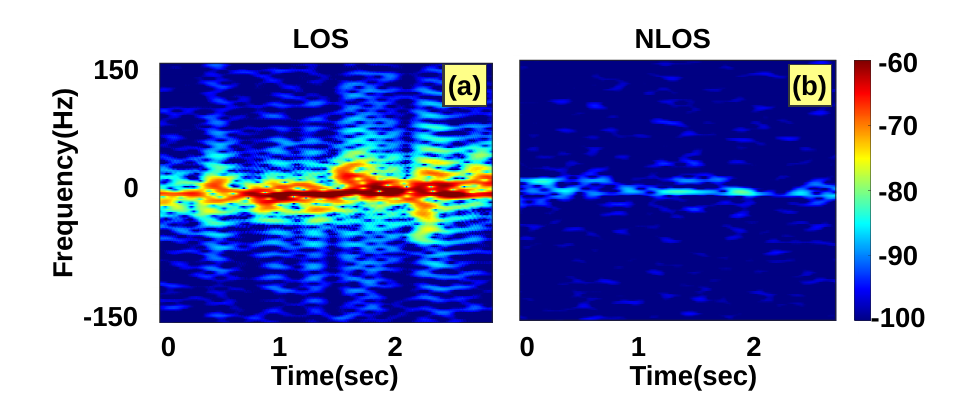}
\caption{Spectrograms of a human walking under different conditions captured using around-the-corner radar at 5.5 GHz: (a) shows trajectory T1 in LOS conditions and (b) shows trajectory T2 in NLOS conditions.}
\label{fig:LOS_NLOS}
\end{figure}

\begin{figure}[htbp]
\centering
\includegraphics[scale=0.4]{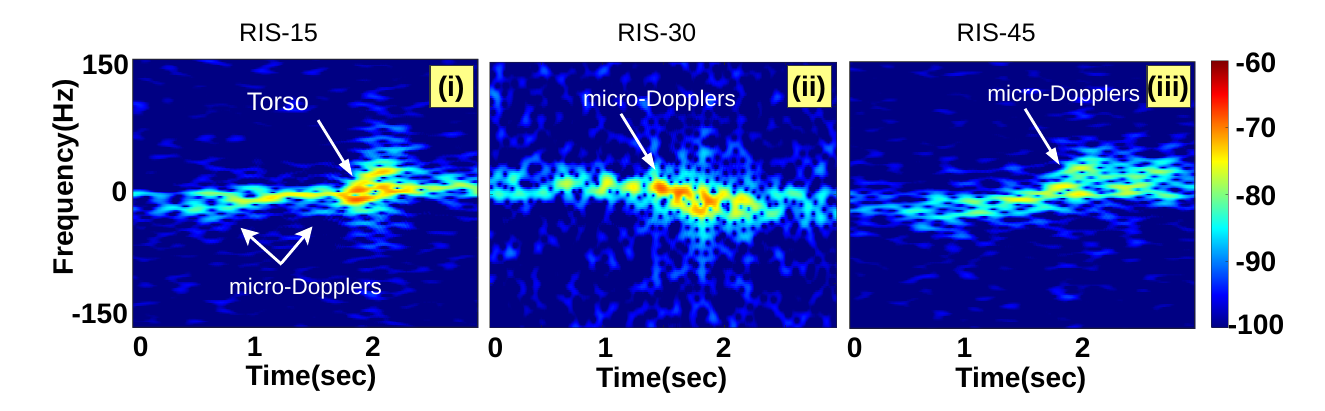}
\caption{Spectrograms of walking motion in NLOS conditions using RIS trajectory T2 for RIS angles of (i) $15^{\circ}$, (ii) $30^{\circ}$, and (iii) $45^{\circ}$, respectively.}
\label{fig:spect_with_RIS}
\end{figure}

 We configured VNA as a monostatic radar to collect measurement data as shown in Fig.\ref{fig:real_exp_setup}a. The VNA performs two-port scattering parameter ($S_{21}$) measurements in the time domain, centered at 5.5 GHz with a bandwidth of 10 kHz. Two HF907 horn antennas are connected to its ports. The VNA transmits at +3 dBm power, and the sampling frequency is set to the maximum limit of 370 Hz to avoid aliasing issues during Doppler measurements.
Each measurement data collection lasts 3 seconds, resulting in approximately 4000 samples. Figure \ref{fig:real_exp_setup}b provides a close-up view of the radar setup.
The micro-Doppler signatures of the human motions in the absence of RIS are depicted in Fig.\ref{fig:LOS_NLOS}. As illustrated in Fig.\ref{fig:LOS_NLOS}a, the human walks tangentially to the radar along trajectory T1 in LOS. Here, we observe that the torso generates strong returns, but the micro-Doppler from the swinging arms and legs is weaker. In a tangential motion, the overall Doppler offsets are low due to the nature of the motion. The target scattered returns from the human following the T2 trajectory under NLOS conditions are shown in Fig.\ref{fig:LOS_NLOS}b. The spectrogram signals are weaker due to radar signals returning via multipath instead of a direct path, as in LOS conditions. The absence of detectable patterns in the NLOS micro-Doppler signatures makes observing human movement around obstacles challenging.
Next, we repeat the measurements after introducing an RIS as described before. In Fig.\ref{fig:spect_with_RIS}, the micro-Doppler signatures for NLOS conditions when RIS is kept at an incident angle of $0^{\circ}$ with respect to the radar are presented. The RIS can switch the beam to different reflection angles. In this experiment, three RIS configurations were considered: RIS-15, RIS-30, and RIS-45, in which the reflected beam is directed toward observation angles of $15^{\circ}$, $30^{\circ}$, and $45^{\circ}$, respectively. In the first row of Fig.\ref{fig:spect_with_RIS}a, the trajectory T1 is depicted where the human is walking along the tangential path to the radar towards the RIS from a distance of 5 m. We observe that the micro-Doppler signatures in NLOS conditions are observed with the incorporation of RIS. The strength of the returns is weaker compared to free space, but it is more robust than the returns in the NLOS case without RIS.

\subsubsection{Measurement Repeatability and Uncertainty Analysis}
To evaluate the reliability of the real-environment $S_{21}$ measurements, a repeatability and uncertainty analysis is performed. Each measurement was captured over a duration of 3~s and was divided into ten non-overlapping windows of 0.3~s, yielding 10 samples per acquisition. Two independent captures were available for the No-RIS, RIS-15, RIS-30, and RIS-45 configurations, resulting in 20 samples for each measurement. After applying STFT, we obtained the spectrograms. For each sample, the mean transmission coefficient and the signal-to-clutter-plus-noise ratio (SCNR) were calculated. The environment-only measurement, where neither RIS nor target motion was present, is used as the clutter-plus-noise reference for SCNR estimation.
The sample mean, variance, root mean squared error (RMSE), 
and two-sided 95\% confidence interval were calculated for each repeated condition ~\cite{montgomery2014applied}. 
The 95\% confidence interval describes uncertainty in the estimated mean; a narrower interval means the mean value is estimated more reliably. The RMSE describes repeatability, where a smaller RMSE means the repeated segment measurements are closer to their mean. 

\begin{figure}[htbp]
  \includegraphics[width=108mm,height=28mm,keepaspectratio]{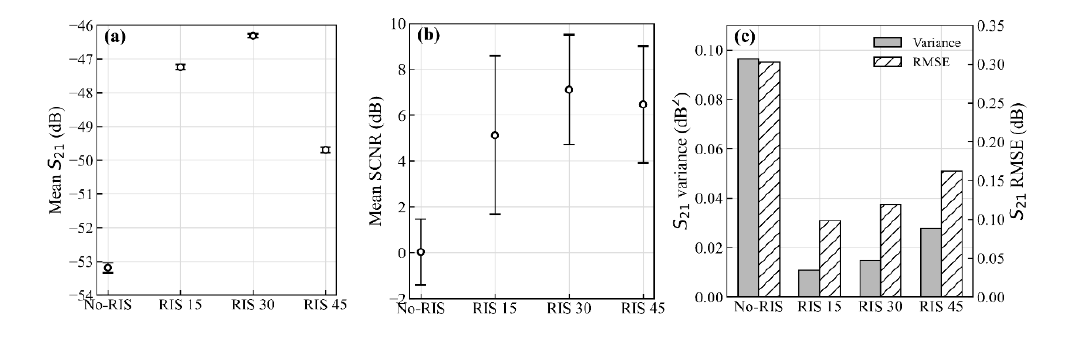}
    \caption{Repeatability and uncertainty analysis for the NLOS reference and RIS-assisted cases: (a) mean $S_{21}$ with 95\% confidence intervals, (b) mean SCNR with 95\% confidence intervals, and (c) $S_{21}$ variance and RMSE across measurement samples.}
    \label{fig:repeatability_combined}
\end{figure}

Fig.~\ref{fig:repeatability_combined}a shows that No-RIS has the weakest mean $S_{21}$. The introduction of RIS improves the mean $S_{21}$ for all cases.RIS-30 gives the strongest received-power improvement.Fig.~\ref{fig:repeatability_combined}b shows that No-RIS has a mean SCNR close to 0~dB, meaning the target response is close to the environment clutter-plus-noise level. RIS-15, RIS-30, and RIS-45 increase the SCNR significantly. 
Fig.~\ref{fig:repeatability_combined}c shows the $S_{21}$ variance and RMSE. RIS cases have lower RMSE than the No-RIS case.  This indicates that the RIS-assisted measurements are more repeatable than the No-RIS case. Overall, the results 
demonstrate that RIS assistance improves both received signal strength and clutter-normalized target response, with RIS-30 giving the strongest performance.
\subsection{ROC-Based Detection Performance Using Measured Environment Data}
To further evaluate radar sensing performance in the real environment, ROC curves are computed directly from the measured data. First, the environment-only measurement is used as the no-target reference. This measurement contains only the environment, static clutter, and noise, and is therefore assigned to the null hypothesis $(H_0)$. Second, the walking-human measurements are used as the target-present data used as the hypothesis, $(H_1)$. These include the No-RIS case and the RIS-assisted cases with different RIS configurations, namely RIS-15, RIS-30, and RIS-45. 
Thus, the environment-only data are used to estimate the false-alarm probability, while the walking-human data are used to estimate the detection probability.
The measured complex channel response is first processed using STFT. The Doppler-time power distribution corresponding to the $i^{th}$ short time duration of the time-domain signal, in the absence (suffix 0) or presence (suffix 1) of target, is expressed as $P_{0/1,i}$. To focus on moving target detection, the detection statistic is computed only within the human micro-Doppler spread, $\Omega_D$, of the Doppler spectrum between 5 and 150 Hz. For each $i^{th}$ Doppler-time sample, a scalar detection statistic is extracted as-
\begin{equation}
D_{0/1,i}= \max_{\Omega_D} \left[P_{0/1,i}-\bar{P}_{\mathrm{env}}\right], 
\end{equation}
where $\bar{P}_{\mathrm{env}}$ is the mean Doppler power obtained from the environment-only measurement over the same Doppler region. 
The detection statistics from the environment-only measurement form the no-target set:  
$\mathcal{D}_0=
\left\{
D_{0,1},D_{0,2},\ldots,D_{0,N_0}
\right\}$ and from the target-present measurements form the target-present set: $
\mathcal{D}_1=
\left\{
D_{1,1},D_{1,2},\ldots,D_{1,N_1}
\right\}.$ Here, $N_0$ and $N_1$ are the numbers of no-target and target-present samples, respectively. 

For a threshold $T_{th}$, the empirical false-alarm probability, $P_{\mathrm{FA}}$, is calculated from the no-target samples as
\begin{equation}
P_{\mathrm{FA}}(T_{th})
=\frac{1}{N_0}\sum_{i=1}^{N_0}\mathbf{1}\left[D_{0,i}>T_{th}\right],
\end{equation}
where $\mathbf{1}[\cdot]$ is the indicator function, which is equal to one when the condition is true and zero otherwise. Similarly, the probability of detection, $P_D$, is calculated from the target-present measurements as
\begin{equation}
P_D(T_{th})=\frac{1}{N_1}\sum_{i=1}^{N_1}\mathbf{1}\left[D_{1,i}>T_{th}\right],
\end{equation}
By changing $T_{th}$, the ROC curve is obtained by plotting $P_D(T_{th})$ against $P_{\mathrm{FA}}(T_{th})$.

\begin{figure}[htbp]
\centering
\includegraphics[scale=0.35]{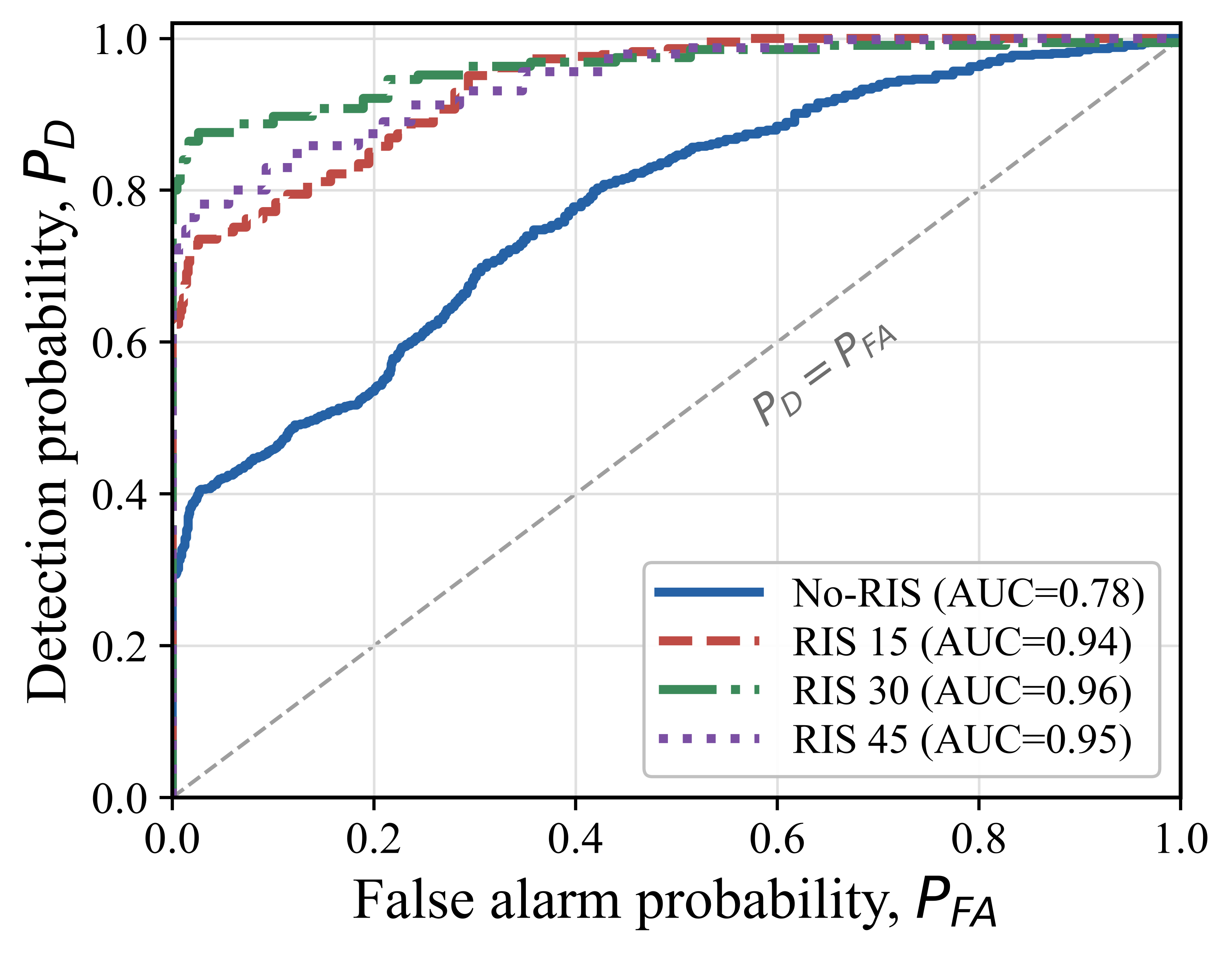}
\caption{Comparison of ROC curves for target detection under no-RIS and RIS-assisted conditions when the target is present.}
\label{fig:roc_measured}
\end{figure}
Fig.~\ref{fig:roc_measured} presents the measured ROC curves obtained from the experimental data. The environment-only measurement is used to compute the false-alarm probability, while the No-RIS and RIS-assisted walking-human measurements are used to compute the detection probability. The ROC curves show that the RIS-assisted cases achieve higher detection probability than the No-RIS case over most of the false-alarm range.

At a low value of $P_{\mathrm{FA}}=0.05$, the measured detection probabilities are 0.42, 0.75, 0.88, and 0.78 for the No-RIS, RIS-15, RIS-30, and RIS-45 cases, respectively. The corresponding area under the ROC curve values are 0.78, 0.94, 0.96, and 0.95. These results indicate that RIS-assisted sensing improves the separation between the target-present measurements and the environment-only reference.
\section{Conclusion and Future Scope}
\label{sec:conclusion}
This work has studied the feasibility and effectiveness of employing a low‑complexity, programmable, quantized RIS for radar target sensing via beam steering. 
We present a comprehensive analysis that hybridizes aperture‑field theory and full‑wave electromagnetic simulations to model both near-field interactions between aperture elements and far-field scattered fields from which the RCS are computed.
Our analysis shows two important insights regarding using RIS for enhancing radar: 1) both the forward and backward RCS must be considered in the two-way path loss model of RIS-enhanced radar, and 2) these two values are not equal for all combinations of $(\theta_i;\theta_d)$. Further, we demonstrate that coarse phase resolution deteriorates the RCS values and overall SNR, while a larger RCS aperture size enhances the bistatic RCS and SNR. A low complexity 1-bit $[16\times10]$ RIS is fabricated and tested in an anechoic chamber and in real radio channels rich with multipath scattering. Our results show that 1-bit RIS shows a considerable reduction in RCS along $\theta_d$ due to grating lobes at $-\theta_d$. Despite these limitations, the simulation and experimental results demonstrate that the 1-bit RIS enhanced the overall radar detection metrics in non-specular regions. 
Additional SCNR, ROC, and repeatability analyses with confidence intervals further confirm the improvement in target detectability and the robustness of the RIS-assisted measurements under multipath conditions.


\begin{thebibliography}{10}

\bibitem{ozdogan2019intelligent}
{\"O}.~{\"O}zdogan, E.~Bj{\"o}rnson, and E.~G. Larsson, ``Intelligent reflecting surfaces: Physics, propagation, and pathloss modeling,'' {\em IEEE Wireless Communications Letters}, vol.~9, no.~5, pp.~581--585, 2019.

\bibitem{liu2024efficient}
Y.~Liu and C.~D. Sarris, ``Efficient computation of scattered fields from reconfigurable intelligent surfaces for propagation modeling,'' {\em IEEE Transactions on Antennas and Propagation}, vol.~72, no.~2, pp.~1817--1826, 2024.

\bibitem{yu2022reconfigurable}
H.~Yu, P.~Li, J.~Su, Z.~Li, S.~Xu, and F.~Yang, ``Reconfigurable bidirectional beam-steering aperture with transmitarray, reflectarray, and transmit-reflect-array modes switching,'' {\em IEEE Transactions on Antennas and Propagation}, vol.~71, no.~1, pp.~581--595, 2022.

\bibitem{huang2019reconfigurable}
C.~Huang, A.~Zappone, G.~C. Alexandropoulos, M.~Debbah, and C.~Yuen, ``Reconfigurable intelligent surfaces for energy efficiency in wireless communication,'' {\em IEEE Transactions on Wireless Communications}, vol.~18, no.~8, pp.~4157--4170, 2019.

\bibitem{10715713}
M.~Deng, M.~Ahmed, A.~Wahid, A.~A. Soofi, W.~U. Khan, F.~Xu, M.~Asif, and Z.~Han, ``Reconfigurable intelligent surfaces enabled vehicular communications: A comprehensive survey of recent advances and future challenges,'' {\em IEEE Transactions on Intelligent Vehicles}, pp.~1--28, 2024.

\bibitem{wu2021intelligent}
Q.~Wu, S.~Zhang, B.~Zheng, C.~You, and R.~Zhang, ``Intelligent reflecting surface-aided wireless communications: A tutorial,'' {\em IEEE Transactions on Communications}, vol.~69, no.~5, pp.~3313--3351, 2021.

\bibitem{elmossallamy2020reconfigurable}
M.~A. ElMossallamy, H.~Zhang, L.~Song, K.~G. Seddik, Z.~Han, and G.~Y. Li, ``Reconfigurable intelligent surfaces for wireless communications: Principles, challenges, and opportunities,'' {\em IEEE Transactions on Cognitive Communications and Networking}, vol.~6, no.~3, pp.~990--1002, 2020.

\bibitem{liu2021reconfigurable}
Y.~Liu, X.~Liu, X.~Mu, T.~Hou, J.~Xu, M.~Di~Renzo, and N.~Al-Dhahir, ``Reconfigurable intelligent surfaces: Principles and opportunities,'' {\em IEEE Communications Surveys \& Tutorials}, vol.~23, no.~3, pp.~1546--1577, 2021.

\bibitem{ram2008through}
S.~S. Ram and H.~Ling, ``Through-wall tracking of human movers using joint doppler and array processing,'' {\em IEEE Geoscience and Remote Sensing Letters}, vol.~5, no.~3, pp.~537--541, 2008.

\bibitem{ram2010simulation}
S.~S. Ram, C.~Christianson, Y.~Kim, and H.~Ling, ``Simulation and analysis of human micro-dopplers in through-wall environments,'' {\em IEEE Transactions on Geoscience and remote sensing}, vol.~48, no.~4, pp.~2015--2023, 2010.

\bibitem{vishwakarma2020micro}
S.~Vishwakarma, A.~Rafiq, and S.~S. Ram, ``Micro-doppler signatures of dynamic humans from around the corner radar,'' in {\em 2020 IEEE International Radar Conference (RADAR)}, pp.~169--174, IEEE, 2020.

\bibitem{dardari2021nlos}
D.~Dardari, N.~Decarli, A.~Guerra, and F.~Guidi, ``Los/nlos near-field localization with a large reconfigurable intelligent surface,'' {\em IEEE Transactions on Wireless Communications}, vol.~21, no.~6, pp.~4282--4294, 2021.

\bibitem{buzzi2021radar}
S.~Buzzi, E.~Grossi, M.~Lops, and L.~Venturino, ``Radar target detection aided by reconfigurable intelligent surfaces,'' {\em IEEE Signal Processing Letters}, vol.~28, pp.~1315--1319, 2021.

\bibitem{buzzi2022foundations}
S.~Buzzi, E.~Grossi, M.~Lops, and L.~Venturino, ``Foundations of mimo radar detection aided by reconfigurable intelligent surfaces,'' {\em IEEE Transactions on Signal Processing}, vol.~70, pp.~1749--1763, 2022.

\bibitem{rihan2022spatial}
M.~Rihan, E.~Grossi, L.~Venturino, and S.~Buzzi, ``Spatial diversity in radar detection via active reconfigurable intelligent surfaces,'' {\em IEEE Signal Processing Letters}, vol.~29, pp.~1242--1246, 2022.

\bibitem{mercuri2023reconfigurable}
M.~Mercuri, E.~Arnieri, R.~De~Marco, P.~Veltri, F.~Crupi, and L.~Boccia, ``Reconfigurable intelligent surface-aided indoor radar monitoring: A feasibility study,'' {\em IEEE Journal of Electromagnetics, RF and Microwaves in Medicine and Biology}, 2023.

\bibitem{yasmeen2024around}
K.~Yasmeen, D.~Kundu, and S.~S. Ram, ``Around-the-corner radar sensing using reconfigurable intelligent surface,'' in {\em 2024 IEEE Microwaves, Antennas, and Propagation Conference (MAPCON)}, pp.~1--4, IEEE, 2024.

\bibitem{patyuchenko2013concept}
A.~Patyuchenko, C.~Tienda, M.~Younis, S.~Bertl, P.~L{\'o}pez-Dekker, and G.~Krieger, ``Concept of a multi-beam reflectarray digital-beam forming synthetic aperture radar,'' in {\em 2013 IEEE International Symposium on Phased Array Systems and Technology}, pp.~346--351, IEEE, 2013.

\bibitem{zhang2023multi}
G.~Zhang, D.~Zhang, Y.~He, J.~Chen, F.~Zhou, and Y.~Chen, ``Multi-person passive wifi indoor localization with intelligent reflecting surface,'' {\em IEEE Transactions on Wireless Communications}, vol.~22, no.~10, pp.~6534--6546, 2023.

\bibitem{wymeersch2020radio}
H.~Wymeersch, J.~He, B.~Denis, A.~Clemente, and M.~Juntti, ``Radio localization and mapping with reconfigurable intelligent surfaces: Challenges, opportunities, and research directions,'' {\em IEEE Vehicular Technology Magazine}, vol.~15, no.~4, pp.~52--61, 2020.

\bibitem{aubry2021reconfigurable}
A.~Aubry, A.~De~Maio, and M.~Rosamilia, ``Reconfigurable intelligent surfaces for n-los radar surveillance,'' {\em IEEE Transactions on Vehicular Technology}, vol.~70, no.~10, pp.~10735--10749, 2021.

\bibitem{chen2023robust}
Z.~Chen, J.~Tang, L.~Huang, Z.-Q. He, K.-K. Wong, and J.~Wang, ``Robust target positioning for reconfigurable intelligent surface assisted mimo radar systems,'' {\em IEEE Transactions on Vehicular Technology}, vol.~72, no.~11, pp.~15098--15102, 2023.

\bibitem{liu2024equivalence}
Y.~Liu, Z.~Liu, S.~V. Hum, and C.~D. Sarris, ``An equivalence principle-based hybrid method for propagation modeling in radio environments with reconfigurable intelligent surfaces,'' {\em IEEE Transactions on Antennas and Propagation}, vol.~72, no.~7, pp.~5961--5973, 2024.

\bibitem{tang2020wireless}
W.~Tang, M.~Z. Chen, X.~Chen, J.~Y. Dai, Y.~Han, M.~Di~Renzo, Y.~Zeng, S.~Jin, Q.~Cheng, and T.~J. Cui, ``Wireless communications with reconfigurable intelligent surface: Path loss modeling and experimental measurement,'' {\em IEEE Transactions on Wireless Communications}, vol.~20, no.~1, pp.~421--439, 2020.

\bibitem{wang2021received}
Z.~Wang, L.~Tan, H.~Yin, K.~Wang, X.~Pei, and D.~Gesbert, ``A received power model for reconfigurable intelligent surface and measurement-based validations,'' in {\em 2021 IEEE 22nd International Workshop on Signal Processing Advances in Wireless Communications (SPAWC)}, pp.~561--565, IEEE, 2021.

\bibitem{sahoo20231}
D.~K. Sahoo, C.~Chakraborty, D.~Kundu, A.~Patnaik, A.~Chakraborty, {\em et~al.}, ``A 1-bit coding reconfigurable metasurface reflector for millimeter wave communications in e-band,'' in {\em 2023 IEEE Microwaves, Antennas, and Propagation Conference (MAPCON)}, pp.~1--4, IEEE, 2023.

\bibitem{jian2022reconfigurable}
M.~Jian, G.~C. Alexandropoulos, E.~Basar, C.~Huang, R.~Liu, Y.~Liu, and C.~Yuen, ``Reconfigurable intelligent surfaces for wireless communications: Overview of hardware designs, channel models, and estimation techniques,'' {\em Intelligent and Converged Networks}, vol.~3, no.~1, pp.~1--32, 2022.

\bibitem{basar2024reconfigurable}
E.~Basar, G.~C. Alexandropoulos, Y.~Liu, Q.~Wu, S.~Jin, C.~Yuen, O.~A. Dobre, and R.~Schober, ``Reconfigurable intelligent surfaces for 6g: Emerging hardware architectures, applications, and open challenges,'' {\em IEEE Vehicular Technology Magazine}, 2024.

\bibitem{khalil2025mitigating}
M.~Khalil, K.~Wang, J.~Lin, and J.~Choi, ``Mitigating phase errors to improve signal quality in ris-assisted satellite communications,'' {\em IEEE Transactions on Vehicular Technology}, 2025.

\bibitem{singh2019controlling}
K.~Singh, M.~U. Afzal, M.~Kovaleva, and K.~P. Esselle, ``Controlling the most significant grating lobes in two-dimensional beam-steering systems with phase-gradient metasurfaces,'' {\em IEEE Transactions on Antennas and Propagation}, vol.~68, no.~3, pp.~1389--1401, 2019.

\bibitem{narayanan2024optimum}
S.~S. Narayanan, U.~K. Khankhoje, and R.~K. Ganti, ``Optimum beamforming and grating lobe mitigation for intelligent reflecting surfaces,'' {\em IEEE Transactions on Antennas and Propagation}, 2024.

\bibitem{xie2024ris}
Z.~Xie, L.~Wu, J.~Zhu, M.~Lops, X.~Huang, and M.~B. Shankar, ``Ris-aided radar for target detection: Clutter region analysis and joint active-passive design,'' {\em IEEE Transactions on Signal Processing}, vol.~72, pp.~1706--1723, 2024.

\bibitem{vejling2025ris}
M.~V. Vejling, H.~Kim, C.~A. Biscio, H.~Wymeersch, and P.~Popovski, ``Ris-assisted high resolution radar sensing,'' {\em IEEE Transactions on Signal Processing}, 2025.

\bibitem{10878502}
H.~Zou, L.~Wu, and Z.~Zhang, ``Target detection for reconfigurable intelligent surface assisted mimo radar,'' {\em IEEE Transactions on Vehicular Technology}, vol.~74, no.~6, pp.~9955--9960, 2025.

\bibitem{yang2025frequency}
H.~Yang, S.~Gong, H.~Liu, C.~Xing, N.~Zhao, and D.~Niyato, ``Frequency diverse array-enabled ris-aided integrated sensing and communication,'' {\em IEEE Transactions on Wireless Communications}, 2025.

\bibitem{balanis2024balanis}
C.~A. Balanis, {\em Balanis' Advanced Engineering Electromagnetics}.
\newblock John Wiley \& Sons, 2024.

\bibitem{willis2005bistatic}
N.~J. Willis, {\em Bistatic radar}, vol.~2.
\newblock SciTech Publishing, 2005.

\bibitem{montgomery2014applied}
D.~C. Montgomery and G.~C. Runger, {\em Applied Statistics and Probability for Engineers}.
\newblock John Wiley \& Sons, 6~ed., 2014.

\end{thebibliography}

\bibliographystyle{ieeetr}
\end{document}